\documentclass[twocolumn,txfonts,showpacs,nofootinbib,superscriptaddress,preprintnumbers,amsmath,amssymb,epsfig,color]{aa}

\usepackage{natbib}
\bibpunct{(}{)}{;}{a}{}{,} 

\usepackage[usenames]{color}
\usepackage{graphicx}

\usepackage{units, subfigure,amssymb,amsmath}
\usepackage[latin1]{inputenc}

\newcommand{\proton}{\ensuremath{^1}\textrm{H}}
\newcommand{\deut}{\ensuremath{^2}\textrm{H}}
\newcommand{\trit}{\ensuremath{^3}\textrm{H}}
\newcommand{\het}{\ensuremath{^3}\textrm{He}}
\newcommand{\hef}{\ensuremath{^4}\textrm{He}}
\newcommand{\deuttohef}{\deut{}/\hef{}}
\newcommand{\hettohef}{\het{}/\hef{}}

\begin{document}
\input epsf
\title{Constraining Galactic cosmic-ray parameters with $Z\leq2$ nuclei}
\author{
       B. Coste\inst{1}
	\and L. Derome\inst{1}
  \and D. Maurin\inst{1}
  \and A. Putze\inst{2}
} 

\offprints{B. Coste, {\tt coste@lpsc.in2p3.fr}}

\institute{
  Laboratoire de Physique Subatomique et de
	Cosmologie, Universit\'e Joseph Fourier Grenoble 1,
  CNRS/IN2P3, Institut Polytechnique de Grenoble,
	53 avenue des Martyrs,
	Grenoble, 38026, France
  \and The Oskar Klein Centre for Cosmoparticle Physics,
  Department of Physics, Stockholm University,
  AlbaNova, SE-10691 Stockholm, Sweden
}

\date{Received / Accepted}

\abstract
{
The secondary-to-primary B/C ratio is widely used to study Galactic cosmic-ray 
propagation processes. The \deuttohef{} and \hettohef{} ratios probe a
different $Z/A$ regime, therefore testing the `universality' of propagation.
}
{
We revisit the constraints on diffusion-model parameters set by the quartet
(\proton{},\deut{},\het{},\hef{}), using the most recent
data as well as updated formulae for the inelastic and production cross-sections.} 
{
The analysis relies on the USINE propagation package
and a Markov Chain Monte Carlo technique to estimate the probability density functions
of the parameters. Simulated data are also used to validate analysis strategies.}
{The fragmentation of CNO cosmic rays (resp.~NeMgSiFe) on the ISM during their propagation
contributes to 20\% (resp.~20\%) of the \deut{} and 15\% (resp.~10\%) of the \het{} flux at
high energy. The C to Fe elements are also responsible for up to 10\% of the \hef{} flux measured
at 1 GeV/n. The analysis of \hettohef{} (and to a less extent \deuttohef{}) data shows
that the transport parameters are consistent with those from the B/C analysis:
the diffusion model with $\delta\sim0.7$ (diffusion slope), $V_c\sim 20$~km~s$^{-1}$ (galactic wind), $V_a\sim 40$~km~s$^{-1}$
(reacceleration) is favoured, but the combination $\delta\sim 0.2$, $V_c\sim 0$, and $V_a\sim 80$~km~s$^{-1}$
is a close second. The confidence intervals on the parameters show that 
the constraints set by the quartet data are competitive with those brought by the B/C data. 
These constraints are tighter when adding the \het{} (or \deut{}) flux measurements, and the tightest when further
adding the He flux. For the latter, the analysis of simulated and real data show an increased sensitivity
to biases.  Using secondary-to-primary ratio along with a loose prior on the
source parameters is recommended to get the most robust constraints on the transport parameters.
}
{
Light nuclei should be systematically considered in the analysis of transport parameters.
They bring independent constraints which are competitive with those obtained from
the B/C analysis.
}

\keywords{Astroparticle physics -- Methods: statistical -- ISM: cosmic rays}

\maketitle


\section{Introduction}

Secondary species in Galactic cosmic rays (GCRs) are produced during the CR journey
from the acceleration sites to the solar neighbourhood, by means of nuclear interactions of
heavier primary species with the interstellar medium. Hence, they are tracers of the CR
transport in the Galaxy \citep[e.g.,][]{2007ARNPS..57..285S}. Studying secondary-to-primary
ratios is useful as it factors out the `unknown' source spectrum of the progenitor,
leaving \deuttohef{}, \hettohef{}, B/C, sub-Fe/Fe|and recently $\bar{p}/p$
\citep{2009A&A...497..991P,2010APh....34..274D}|suitable quantities to constrain the
transport parameters for species $Z\leq30$.

Most secondary-to-primary ratios have  $A/Z\sim 2$, and in that respect,
\hettohef{} is unique since it probes a different regime and allows to address the issue
of the  `universality' of propagation histories. For instance, in an analysis in the
leaky-box model (LBM) framework, \citet{1997AdSpR..19..755W} found that \hettohef{} data
imply a similar propagation history for the light and heavier species (which was disputed
in earlier papers). Webber also argued that the situation with regard to the \deuttohef{}
ratio is less clear, because the uncertainties on the measurements are large (mainly due to
instrumental and atmospheric corrections). H and He spectra are the most abundant
species in the cosmic radiation, and thus \deut{} and \het{} are the most abundant secondary
species in GCRs. However, achieving a good mass resolution|especially at high energy|is
experimentally challenging. This explains why the elemental B/C ratio received more focus
both experimentally and theoretically (thanks to its higher precision data w.r.t. to the
quartet data).

From the modelling side, after the first thorough and pioneering studies performed in the
60's-70's 
\citep{1963PThPh..30..615B,1969ApJ...155..587R,Meyer72,1972Ap&SS..17..186M,1974Ap&SS..30..187R,1976ApJ...206..616M},
the interest for the quartet nuclei somewhat stalled. Several updated analyses of the
propagation parameters from the quartet were published as new data became available (see
Table~\ref{tab:refs_quartet} for references). However, very few dedicated studies were
carried out in the 80's \citep{1986ApJ...311..425B,1987ApJ...312..178W}, likewise in the
90's \citep{1990ICRC....3..404W,1994ApJ...431..705S,1997AdSpR..19..755W}, and none in the
00's.  This is certainly related to the very slow pace at which new data became available
in this period. Curiously, the most recent published data have not really been properly
interpreted, i.e. for \deuttohef{} data, IMAX92 \citep{2000AIPC..528..425D} and
AMS-01~\citep{2011ApJ...736..105A}; and for
\hettohef{} data, IMAX92 \citep{2000ApJ...533..281M}, SMILI-II
\citep{2000ApJ...534..757A}, AMS-01 \citep{Xiong:2003hp}, BESS98
\citep{2003ICRC....4.1805M}, CAPRICE98 \citep{2003ICRC....4.1809M}. Furthermore, almost all
analyses have been performed in the successful but simplistic LBM, but in a few
studies\footnote{\citet{1994ApJ...431..705S} used a 1D diffusion model with reacceleration
whereas \citet{1997AdSpR..19..817W} relied on a Monte Carlo calculation; both studies
conclude similarly (consistency with the grammage required for heavier species to produce
the light secondaries). A preliminary effort based on the GALPROP propagation code was
also carried out in \citet{2003ICRC....4.1917M}.}. At the same time, the analysis of the
B/C ratio has been scrutinised in more details. For instance, to replace the old usage of
matching the data by means of an inefficient manual scan of the parameter space
\citep[e.g.,][]{2001ApJ...547..264J}, more systematic scans were carried out (on the B/C
and sub-Fe/Fe ratio) to get best-fit values as well as uncertainties on the parameters
\citep{2001ApJ...555..585M,2005JCAP...09..010L,2008JCAP...10..018E,2010APh....34..274D}.
A recent improvement is the use of Markov Chain Monte Carlo (MCMC) techniques  to
directly access the probability-density function (PDF) of the GCR transport and source
parameters
\citep{2009A&A...497..991P,2010A&A...516A..66P,2011A&A...526A.101P,2011ApJ...729..106T}. 

In this paper, we revisit the constraints set by the quartet nuclei and their
consistency with the results of heavier nuclei. In the context of the forthcoming PAMELA
and AMS-02 data on these ratios, we also discuss the strategy to adopt and intrinsic
limitations of the transport parameters reconstruction. For that purpose, we take advantage
of the data taken in the last decade as well as simulated data of any precision, and analyse them
with an MCMC technique implemented in the USINE propagation code. This extends and complements
analyses of the B/C and primary nuclei \citep{2010A&A...516A..66P,2011A&A...526A.101P} 
in a 1D diffusion model. 

The paper is organised as follows. In Sect.~\ref{sec:Model}, we briefly recall the main
ingredients of the 1D diffusion model and the MCMC analysis. We also list the parameters
which are constrained. The simulated data and their analysis are described in Sect.~\ref{sec:fake}.
The analysis of the real data is given in Sect.~\ref{sec:res}. We conclude in Sect.~\ref{sec:conclusion}. 
Appendix~\ref{app:quartet} gathers the data sets and the updated cross-sections
used in the quartet analysis.

\section{MCMC technique, propagation and parameters\label{sec:Model}}

The MCMC technique and its use in the USINE propagation code is
detailed in \citet{2009A&A...497..991P} and summarised in \citet{2010A&A...516A..66P}. The
full details regarding the 1D transport model can be found in \citet{2010A&A...516A..66P}.
Below, we only provide a brief description.

\subsection{An MCMC technique for the PDF of the parameters}

The MCMC method, based on Bayesian statistics, is used to estimate the 
full distribution (conditional PDF) given
some experimental data and some prior density for these parameters. 
Our chains are based on the Metropolis-Hastings algorithm, which
ensures that the distribution of the chain asymptotically
tends to the target PDF. 

The chain analysis refers to the selection of a subset of points from the chains
(to get a reliable estimate of the PDF). The steps at the beginning of the chain are
 discarded (burn-in length) if they are too far of the region of interest.
 Sets of independent samples are obtained by thinning the
chain (over the correlation length).
The final results of the MCMC analysis are the joint and marginalised PDFs.
They are obtained by counting the number of samples within the related
region of the parameter space.

\subsection{1D Propagation model and parameters}

The Galaxy is modelled to be an infinite thin disc of half-thickness $h$, which contains the gas and
the sources of CRs. The diffusive halo region (where the gas density is assumed to be equal to 0)
extends to $+L$ and $-L$ above and below the disc. A constant wind
$\vec{V}(\vec{r})={\rm sign}(z) \cdot V_c \times \vec{e}_z$, perpendicular to the Galactic
plane, is assumed. In this framework, CRs diffuse in the disc and in the halo
independently of their position. Such semi-analytical models are faster than full numerical
codes (GALPROP\footnote{\tt
http://galprop.stanford.edu/} and DRAGON\footnote{\tt
http://www.desy.de/$\sim$maccione/DRAGON/}), which is an advantage for sampling techniques like MCMC approaches.

\subsubsection{Transport equation}

The differential density $N^j$ of the nucleus $j$ is a function of the total energy $E$ and
the position $\vec{r}$ in the Galaxy.  Assuming a steady state, the transport equation can be
written in a compact form as 
\begin{equation}
{\cal L}^j N^j + \frac{\partial}{\partial E}\left( b^j N^j - c^j \frac{\partial N^j}{\partial E} \right) = {\cal S}^j\;.
\label{eq:CR}
\end{equation}
The operator ${\cal L}$ (we omit the superscript $j$) describes the diffusion
$K(\vec{r},E)$ and the convection $\vec{V}(\vec{r})$ in the Galaxy, but also the decay rate
$\Gamma_{\rm rad}(E)= 1/(\gamma\tau_0)$ if the nucleus is radioactive, and the destruction
rate $\Gamma_{\rm inel}(\vec{r},E)=\sum_{ISM} n_{\rm ISM}(\vec{r}) v \sigma_{\rm inel}(E)$ for
collisions with the interstellar matter (ISM), in the form
\begin{equation}
{\cal L}(\vec{r},E) =  -\vec{\nabla} \cdot (K\vec{\nabla}) + \vec{\nabla}\cdot\vec{V} +
     \Gamma_{\rm rad} + \Gamma_{\rm inel}.
\label{eq:operator}
\end{equation}

The coefficients $b$ and $c$ in Eq.~(\ref{eq:CR}) are respectively first and
second order gains/losses in energy, with
\begin{eqnarray}
\label{eq:b}
b\,(\vec{r},E)&=& \big\langle\frac{dE}{dt}\big\rangle_{\rm ion,\,coul.} 
   - \frac{\vec{\nabla}.\vec{V}}{3} E_k\left(\frac{2m+E_k}{m+E_k}\right)
	 \\\nonumber
   &  & + \;\; \frac{(1+\beta^2)}{E} \times K_{\rm pp},\\
\label{eq:c}
c\,(\vec{r},E)&=&  \beta^2 \times K_{\rm pp}.
\end{eqnarray}
In Eq.~(\ref{eq:b}), the ionisation and Coulomb energy losses are taken from
\citet{1994A&A...286..983M} and \citet{1998ApJ...509..212S}. The divergence of the Galactic
wind $\vec{V}$ gives rise to an energy loss term related to the adiabatic expansion of
cosmic rays. The last term is a first order contribution in energy from reacceleration.
Equation~(\ref{eq:c}) corresponds to a diffusion in momentum space, leading to an energy gain.
The associated diffusion coefficient
$K_{\rm pp}$ (in momentum space) is taken from the model of minimal reacceleration by the
interstellar turbulence \citep{1988SvAL...14..132O,1994ApJ...431..705S}. It is related to the
spatial diffusion coefficient $K$ by
\begin{equation}
K_{\rm pp}\times K= \frac{4}{3}\;V_a^2\;\frac{p^2}{\delta\,(4-\delta^2)\,(4-\delta)},
\label{eq:Va}
\end{equation}
where $V_a$ is the Alfv\'enic speed in the medium.

We refer the reader to App.~A of \citet{2010A&A...516A..66P}
for the solution to Eq.~(\ref{eq:CR}) in the 1D geometry.

\subsubsection{Free parameters of the analysis}

The exact energy dependence of the source and transport parameters is unknown, but they
are expected to be power laws of ${\cal R}=pc/Ze$ (rigidity of the particle). 

The low-energy diffusion coefficient requires a $\beta=v/c$ factor that takes into account the
inevitable effect of particle velocity on the diffusion rate. However, the recent analysis
of the turbulence dissipation effects on the transport coefficient has shown that this
coefficient could increase at low-energy \citep{2006ApJ...642..902P,2010ApJ...725.2110S}.
Following \citet{2010A&A...516A..67M}, it is parametrised to be
\begin{equation}
K(E)= \beta^{\eta_T} \cdot K_0 {\cal R}^\delta \;.
\label{eq:eta_T}
\end{equation}
The default value used for this analysis is $\eta_T=1$.
The two other transport parameters are $V_c$, the constant convective wind perpendicular to the disc,
and $V_a$, the Alfv\'enic speed regulating the reacceleration strength [see Eq.~(\ref{eq:Va})].
The two models considered in this paper are given in Table \ref{tab:class_models}.
\begin{table}[!t]
\caption{Models tested in the paper.}
\label{tab:class_models}
\centering
\begin{tabular}{ccc} \hline\hline
Model    & $\!\!\!$Transport parameters$\!\!\!$  & Description  \\
\hline\vspace{-0.2cm}
& \multicolumn{2}{c}{} \\ 
II       & $\{K_0,\, \delta, \, V_a\}$         & Diffusion + reacceleration \vspace{0.05cm}\\
III      & $\{K_0,\, \delta, \, V_c, \, V_a\}$ & Diff. + conv. + reac.    \vspace{0.cm}\\
\hline
\end{tabular}
\note{\tiny For the sake of consistency, the model identification follows that of
\citet{2009A&A...497..991P,2010A&A...516A..66P,2011A&A...526A.101P}
and \citet{2010A&A...516A..67M}.
}
\end{table}

The low-energy primary source spectrum from acceleration models (e.g.,
\citealt{1983RPPh...46..973D,1994ApJS...90..561J}) is also unknown.
We parametrise it to be
\begin{equation}
        Q_{E_{k/n}}(E) \equiv \frac{dQ}{dE_{k/n}}  
        = q \cdot \beta^{\eta_S} \cdot {\cal R}^{- \alpha},
       \label{eq:source_spec}
    \end{equation}
where $q$ is the normalisation. The reference low-energy shape corresponds
to $\eta_S=-1$ (to have $dQ/dp\propto p^{-\alpha}$, i.e. a pure power-law).

The halo size of the Galaxy $L$ cannot be solely determined from secondary-to-primary
stable ratios and requires a radioactive species to lift the degeneracy between $K_0$ and
$L$. However, the range of allowed values is still very loosely
constrained~\citep[e.g.,][]{2010A&A...516A..66P}. As the transport and source parameters
can always be rescaled would a different choice of $L$ assumed (see the scaling relations
given in \citealt{2010A&A...516A..67M}, where $\delta$ is shown not to depend on
$L$), we fix it to $L=4$~kpc. This will also ease the comparison of the results obtained
in this paper with those of our previous studies
\citep{2010A&A...516A..66P,2011A&A...526A.101P}.

\section{MCMC analysis on artificial data sets\label{sec:fake}}

MCMC techniques make the scan of high-dimensional parameter spaces possible, such that a simultaneously estimation of transport and source parameters  is possible \citep{2009A&A...497..991P}. However, transport parameters are shown to be strongly degenerated for the B/C ratio data in the range $0.1-100$ GeV/nuc \citep{2010A&A...516A..67M}, and source and transport parameters are correlated \citep{2009A&A...497..991P,2010A&A...516A..66P}. For GCR data in general, the fact that primary fluxes and secondary fluxes are not measured to the same accuracy\footnote{Statistical uncertainties are smaller for primary fluxes (more abundant than secondary fluxes), but the latter a more prone to systematics than ratios (e.g. secondary-to-primary ratios used to fit transport parameters).} can bias or prevent an accurate determination of these parameters: a simultaneous fit has been observed to be driven by the more accurately measured primary flux \citep{2011A&A...526A.101P}. This, although statistically correct, might not maximise the information obtained on the transport parameters. Therefore, several strategies can be considered when dealing with GCR data:
\begin{itemize}
\item a combined analysis of secondary-to-primary ratio and primary flux to constrain simultaneously the source and transport parameters;
\item a secondary-to-primary ratio analysis only, either fixing the source parameters (i.e., using a strong prior), or using a loose prior.
\item a primary flux analysis only, either fixing the transport parameters (i.e., using a strong prior), or using a loose prior.
\end{itemize}
In the literature, the strong prior approach has almost always been used to determine the transport or the source parameters. The issue we wish to address is how sensitive the sought parameters are to various strategies. This is the motivation to introduce artificial data, i.e. an ideal case study, as opposed to the case of real data where several other complications can arise (systematics in the data and/or the use of the incorrect propagation model or solar modulation model/level).     


\subsection{Sets of artificial data} 
\begin{table*}[!t]
\caption{Simulated data analysis for several {\em Models} (input parameters in {\em italic}) with $L=4$~kpc: each line corresponds to the
MCMC-reconstructed values (most-probable value, and relative uncertainties corresponding to 
the 68\% CI) based on a given data/parameters option (see Sect.~\ref{sec:strategy}). The last column gives the value of
the best $\chi^2$/d.o.f. configuration found (corresponding to the curves shown in Fig.~\ref{fig:ModelVc=0}).}
\label{tab:simu_param}\centering
\begin{tabular}{lcccccccc} \hline\hline
$\!\!\!$Option: data/params$\!\!\!\!\!$ &   $\eta_{T}$ & $K_{0}\times 10^2$    &   $\delta$  &     $V_{c}$   &   $V_a$         &  $\alpha$  &  $\eta_{S}$  & $\!\!\!\chi^2_{\rm best}/$d.o.f.$\!\!\!$ \\
                              &      -       & $\!\!\!$(kpc$^2$\,Myr$^{-1})\!\!\!$ &      -      &(km\,s$^{-1})$ &  (km\,s$^{-1})$ &    -       &      -       &    -   \\
\hline
\multicolumn{9}{c}{\vspace{-0.2cm}} \\
\multicolumn{9}{c}{\em Model II\vspace{0.05cm}} \\
                             & {\em  1 }                    &  {\em 10.0}                   &  {\em 0.2}                  &   {\em \dots}              &  {\em 70 }                  &  {\em 2.3}                  &{\em 1}                      & {\em \dots} \vspace{0.1cm}\\
1: \het{}/\hef{}+He$^{10\%}$ &           [1]                & $10.3^{+  3.9\%}_{ -3.9\%}$   & $0.185^{+ 54\%}_{ -5.4\%}$  &        \dots               & $72.7^{+  6.3\%}_{ -5.6\%}$ & $2.29^{+  0.9\%}_{ -1.3\%}$ & $0.78^{+ 13.\%}_{-17.\%}$   & 0.93 \vspace{0.1cm} \\
2: \het{}/\hef{}+He$^{1\%}$  &           [1]                & $10.2^{+  2.0\%}_{ -3.9\%}$   & $0.192^{+ 3.1\%}_{ -2.1\%}$ &        \dots               & $73.0^{+  2.7\%}_{ -4.1\%}$ & $2.29^{+  0.3\%}_{ -0.3\%}$ & $0.88^{+  3.4\%}_{ -5.7\%}$ & 1.08 \vspace{0.1cm} \\
2':    2 + src=true          &           [1]                & $9.7^{+  3.1\%}_{ -3.1\%}$    & $0.199^{+ 3.0\%}_{ -4.0\%}$ &        \dots               & $68.0^{+  4.4\%}_{ -2.9\%}$ &         [2.3]                &            [1]               & 0.97 \vspace{0.1cm} \\
3: \het{}/\hef{}             &           [1]                & $11.5^{+   13.\%}_{-  23.\%}$ & $0.19^{+   21\%}_{ -16\%}$  &        \dots               & $39.4^{+  68.\%}_{ -53.\%}$ & $2.70^{+148\%}_{-85\%}$     & $1.5^{+326\%}_{-106\%}$     & 0.87 \vspace{0.1cm} \\
3':    3 + src=true          &           [1]                & $10.1^{+  3.0\%}_{ -4.0\%}$   & $0.196^{+ 5.1\%}_{ -7.1\%}$ &        \dots               & $69.6^{+  5.9\%}_{ -6.6\%}$ &         [2.3]                &            [1]               & 0.88 \vspace{0.1cm} \\
4: \het{}/\hef{} + src=prior &           [1]                & $9.0^{+ 11.\%}_{-11.\%}$      & $0.2^{+   15.\%}_{-10. \%}$ &        \dots               & $73.3^{+  5.5\%}_{ -8.2\%}$ & $[1.8,2.5]      $                  & $[-2,+2]      $ & 1.08 \vspace{0.1cm} \\
\hline
\multicolumn{9}{c}{\vspace{-0.2cm}} \\
\multicolumn{9}{c}{\em Model III\vspace{0.05cm}} \\
                             &  {\em 1.5}                  &  {\em 0.75}                   &  {\em 0.7}                  &  {\em 18 }                 & {\em 41 }                   &  {\em 2.3}                  &  {\em 1  }                  & {\em \dots} \vspace{0.1cm}\\
1: \het{}/\hef{}+He$^{10\%}$ & $1.66^{+ 42.\%}_{-13.\%}$   & $1.5^{+ 40.\%}_{-47.\%}$      & $0.51^{+ 33.\%}_{ -7.8\%}$  & $18.4^{+ 17.\%}_{-10.\%}$  & $54.1^{+ 18.\%}_{ -8.9\%}$  & $2.29^{+ 0.9\%}_{-1.7\%}$   & $1.^{+  7.0\%}_{-36.\%}$    & 1.43 \vspace{0.1cm} \\
2: \het{}/\hef{}+He$^{1\%}$  & $1.42^{+  2.8\%}_{ -1.4\%}$ & $0.62^{+  9.7\%}_{ -9.7\%}$   & $0.725^{+  0.8\%}_{ -3.7\%}$& $19.7^{+  3.6\%}_{ -4.1\%}$& $37.1^{+  9.2\%}_{ -1.3\%}$ & $2.334^{+ 0.3\%}_{-1.1\%}$  & $0.98^{+  4.1\%}_{ -2.0\%}$ & 1.00 \vspace{0.1cm} \\
2':    2 + src=true          & $1.48^{+  0.7\%}_{ -0.7\%}$ & $0.7^{+  4.3\%}_{ -5.7\%}$    & $0.71^{+  1.4\%}_{ -1.4\%}$ & $18.3^{+  1.6\%}_{ -3.3\%}$& $40.7^{+  2.7\%}_{ -2.2\%}$ &         [2.3]                &            [1]               & 1.01     \vspace{0.1cm} \\
3: \het{}/\hef{}             & $1.47^{+  8.8\%}_{-12.\%}$  & $0.92^{+ 89.\%}_{-48.\%}$     & $0.57^{+ 21.\%}_{-14.\%}$   & $20.5^{+ 18.\%}_{-14.\%}$  & $58.0^{+ 22.\%}_{-15.\%}$   & $0.12^{+708\%}_{-92.\%}$    & $-2.2^{+50.\%}_{-36.\%}$    & 0.83 \vspace{0.1cm} \\
3':    3 + src=true          & $1.34^{+ 29.\%}_{ -0.7\%}$  & $0.48^{+139\%}_{-60.\%}$      & $0.68^{+ 24.\%}_{-22.\%}$   & $18.1^{+  6.1\%}_{ -3.3\%}$& $44.8^{+ 14.\%}_{-28.\%}$   &         [2.3]                &            [1]               & 0.89 \vspace{0.1cm} \\
4: \het{}/\hef{} + src=prior & $1.38^{+ 13.\%}_{-16.\%}$   & $0.37^{+132\%}_{-62.\%}$      & $0.65^{+ 29.\%}_{ -1.7\%}$  & $20.3^{+ 13.\%}_{ -4.0\%}$ & $42.2^{+ 15.\%}_{-25.\%}$   & $[1.8,2.5]      $                  & $[-2,+2]      $ & 0.96 \vspace{0.1cm} \\
\hline
\multicolumn{9}{c}{\vspace{-0.2cm}} \\
\multicolumn{9}{c}{\em Model III: analysis with Model II \vspace{0.05cm}} \\
                             &  {\em 1.5}                  &  {\em 0.75}                   &  {\em 0.7}                  &  {\em 18 }                 & {\em 41 }                   &  {\em 2.3}                  &  {\em 1  }                  & {\em \dots} \vspace{0.1cm}\\
1: \het{}/\hef{}+He$^{10\%}$ &           [1]                & $13.8^{+  3.6\%}_{ -5.1\%}$   & $0.21^{+  4.8\%}_{ -4.8\%}$ &        \dots               & $126^{+  4.8\%}_{ -4.0\%}$  & $2.3^{+  0.9\%}_{ -1.7\%}$  & $0.24^{+ 42.\%}_{-29.\%}$   & 4.0  \vspace{0.1cm} \\
2: \het{}/\hef{}+He$^{1\%}$  &           [1]                & $11.4^{+  3.5\%}_{ -1.8\%}$   & $0.263^{+  3.8\%}_{ -3.8\%}$&        \dots               & $85^{+  2.4\%}_{ -3.5\%}$   & $2.4^{+  0.4\%}_{ -0.4\%}$  & $1.19^{+  1.7\%}_{ -3.4\%}$ & 18   \vspace{0.1cm} \\
4: \het{}/\hef{} + src=prior &           [1]                & $20.7^{+  20.\%}_{ -8.2\%}$   & $0.089^{+ 31.\%}_{-18.\%}$  &        \dots               & $107^{+  5.6\%}_{-12.\%}$   & $[1.8,2.5]      $                  & $[-2,+2]      $ & 2.1  \vspace{0.1cm} \\ 
\hline
\multicolumn{9}{c}{\vspace{-0.2cm}} \\
\multicolumn{9}{c}{\em  Model II: analysis with Model III \vspace{0.05cm}} \\
                             & {\em  1 }                   &  {\em 10.0}                   &  {\em 0.2}                  &   {\em  0   }              &  {\em 70 }                  &  {\em 2.3}                  &{\em 1}                      & {\em \dots} \vspace{0.1cm}\\
1: \het{}/\hef{}+He$^{10\%}$ & $ 1.37^{+ 5.1\%}_{-4.4\%}$  & $ 4.0^{+ 65\%}_{-45\%}$       & $ 0.29^{+ 24\%}_{-21\%}$        & $17.2^{+  9.3\%}_{-41\%}$& $73.5^{+  7.2\%}_{-17\%}$     & $2.21^{+ 1.8\%}_{-0.5\%}$     & $ 0.80^{+ 12\%}_{-29\%}$     & 0.97 \vspace{0.1cm} \\
2: \het{}/\hef{}+He$^{1\%}$  & $ 0.87^{+ 9.2\%}_{-5.7\%}$  & $ 7.4^{+ 19\%}_{-27\%}$       & $ 0.25^{+ 20\%}_{-8.0\%}$       & $ 8.6^{+ 62\%}_{- 50\%}$& $76.3^{+  3.5\%}_{ -4.8\%}$   & $2.23^{+ 0.4\%}_{-0.4\%}$     & $ 0.68^{+7.4\%}_{-10\%}$     & 1.1  \vspace{0.1cm} \\
4: \het{}/\hef{} + src=prior & $ 0.43^{+193\%}_{-107\%}$   & $ 4.9^{+ 49\%}_{-73\%}$       & $ 0.24^{+ 50\%}_{-21\%}$        & $18.4^{+ 12\%}_{-59\%}$  & $69.9^{+ 12\%}_{-18\%}$       & $[1.8,2.5]      $                  & $[-2,+2]      $ & 1.05 \vspace{0.1cm} \\
\hline
\end{tabular}
\note{\tiny A value in square brackets corresponds to the fixed value of the parameter for the analysis. An interval in square brackets
corresponds to the prior used for the analysis (the posterior PDF obtained is close to the prior).}
\end{table*}

To be as realistic as possible, we choose models that roughly reproduce the actual data points (see Fig.~\ref{fig:envelopes}), but also match the typical energy coverage, number of data points, central value and spread (error bars) of the measurements\footnote{The uncertainty on the H and He fluxes is a few percents (for the recent PAMELA data, \citealt{2011Sci...332...69A}) and several tens of percents for the \hettohef{} ratio.}. To speed-up the calculation and for this section only, we assume that all \het{} comes from \hef{} (see Sect.~\ref{sec:frac_contrib} for all the relevant progenitors). No systematic errors were added although they may set a fundamental limitation in recovering the cosmic-ray parameters. 
In practice, the statistical errors for the artificial data sets correspond to the sigma of the standard Gaussian deviations used to randomise the data points around their model value: \hettohef{} was generated with statistical errors of 10\% while He fluxes were generated with 1\% and 10\% errors, to simulate the situation where primary fluxes are `more accurately' or `equally' measured (in terms of statistics) than the secondary-to-primary ratio.

The parameters of the two models used to simulate the data are listed in the two {\em italic} lines in Table~\ref{tab:simu_param}, denoted {\em Model~II} and {\em Model~III}. They correspond to extreme values of the diffusion slope $\delta$, but which still roughly fall in the range of values found for instance from the B/C analysis \citep{2010A&A...516A..66P}: for {\em Model II} with reacceleration only ($V_c=0$), $\delta$ is generally found to fall between $0.1$ and $0.3$, whereas for {\em Model III} with convection and reacceleration, $\delta$ is generally found to fall into the $0.6-0.8$ range \citep{2001ApJ...547..264J,2010A&A...516A..67M}.

\subsection{Strategies to analyse the data\label{sec:strategy}}

To test the impact on the reconstruction of the {\em transport} ($\eta_T$, $K_0$, $\delta$, $V_a$, and $V_c$) and/or {\em source}
($\alpha$ and $\eta_S$) parameters, we test the following combinations (data set$|$model parameters) for the analysis.
\paragraph{\hettohef{} + He data}
  \begin{description}
    \item[Option~1 ($\sigma_{\rm He} \!=\! 10 \%$):] {\em transport} + {\em source};
    \item[Option~2 ($\sigma_{\rm He} \!=\! 1 \%$):] {\em transport} + {\em source};
    \item[Option~2' ($\sigma_{\rm He} \!=\! 1 \%$):] {\em transport} (source = `true' value);
  \end{description}
\paragraph{\hettohef{} ratio only}
  \begin{description}
    \item[Option~3:] {\em transport} + {\em source};
    \item[Option~3':] {\em transport} (source = `true' value);
    \item[Option~4:] {\em transport} (source = weak prior).
  \end{description}

We find that the He data alone cannot constrain the transport parameters (not shown here), in agreement
with \citet{2011A&A...526A.101P} results (strong degeneracy between $\alpha$
and $\delta$, but also with $K_0$, $V_a$, and $\eta_T$).

\subsection{Analysis of the artificial data}

\begin{figure}[!t] 
\centering
\includegraphics[width=\columnwidth]{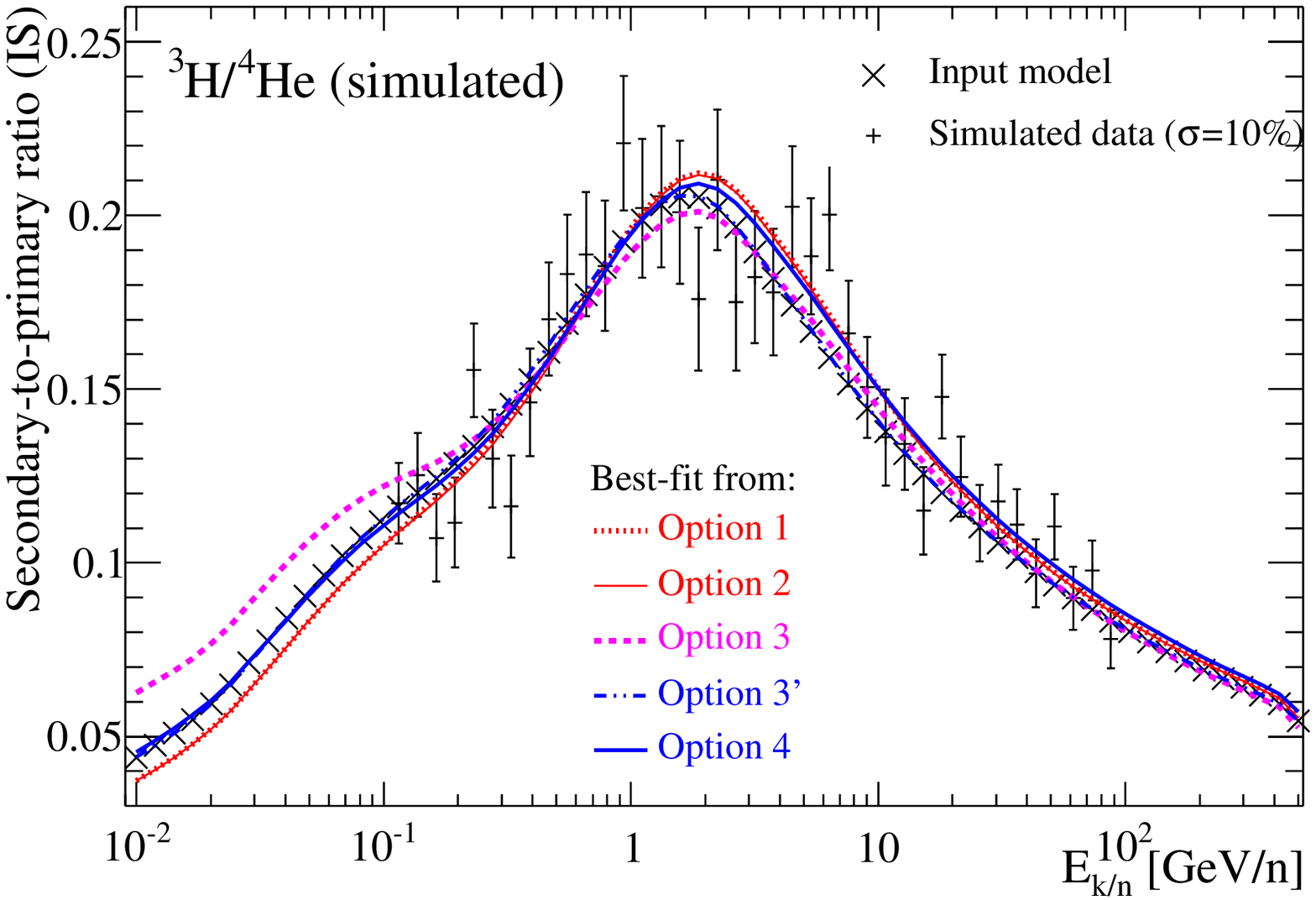}
\includegraphics[width=\columnwidth]{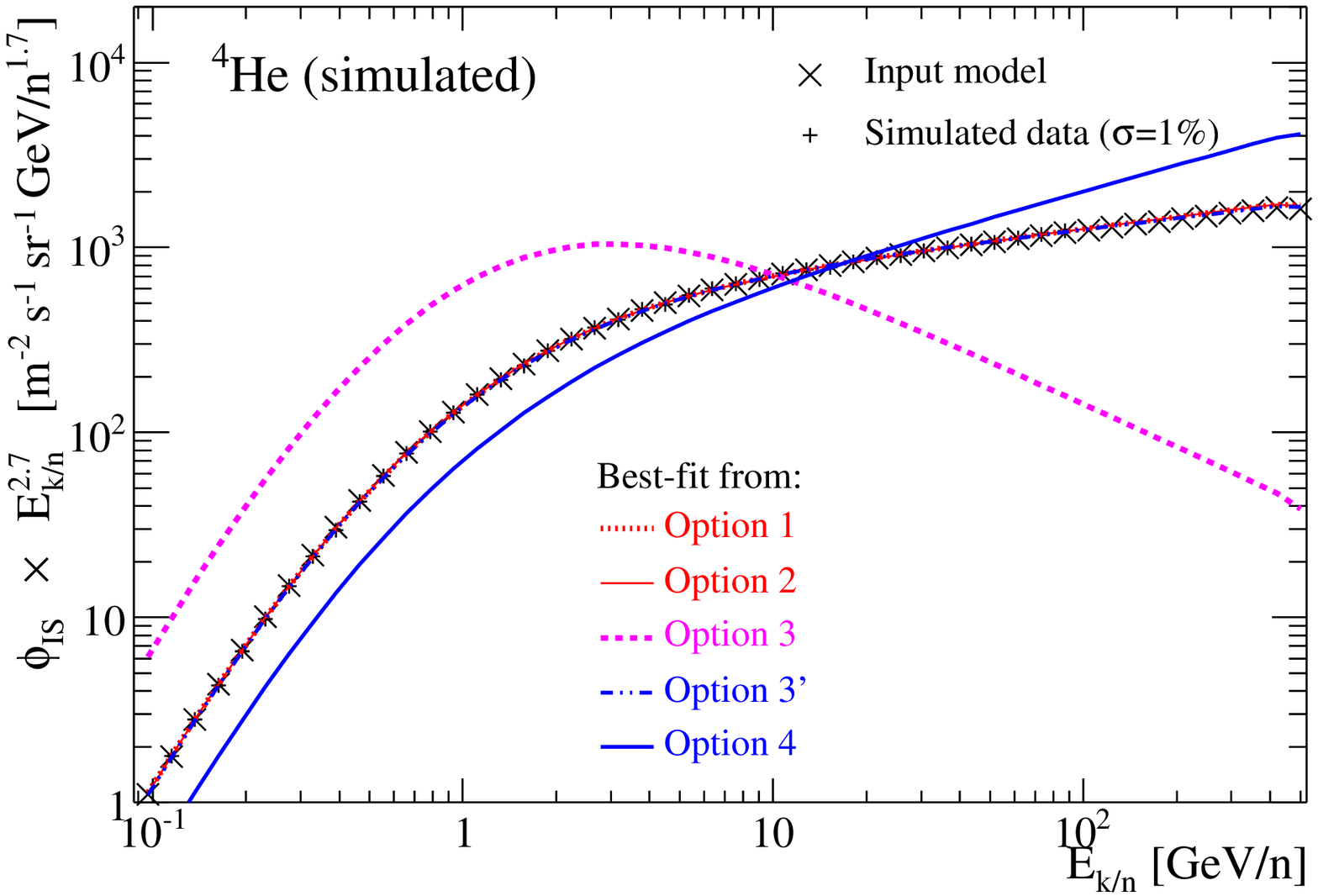}
\caption{Analysis of simulated interstellar (IS) data sets for the \hettohef{}
ratio (top panel) and the \hef{} flux times $E_{k/n}^{2.7}$ (bottom panel) based on Model II
($\times$ symbols), the parameters of which are given in  Table~\ref{tab:simu_param}).
The best-fit reconstructed curves correspond to the different `options' 
given in Sect.~\ref{sec:strategy}.}
\label{fig:ModelVc=0}
\end{figure}

In a first step, we used the MCMC technique to estimate the best-fit parameters.
The \hettohef{} ratio and \hef{} flux are shown for model II (crosses) and the corresponding simulated data (plusses) in Fig.~\ref{fig:ModelVc=0}.
When both the \hettohef{} and \hef{} data are
included in the fit  (options 1 and 2, red dotted and red solid lines), the initial
flux (crosses) is perfectly recovered for \hef{}, and very well recovered for \hettohef{}.
When the fit is only based on \hettohef{} (options 3 and 5, magenta dashed and blue solid
lines), the initial flux is obviously not recovered (unless the source parameters are set to
the true value as in option~3'), but the \hettohef{} ratio is consistent with the data.
Unsurprisingly, the associated $\chi^2_{\rm best}$/d.o.f. values (last column of
Table~\ref{tab:simu_param}) are close to 1.

\begin{figure}[!t]
  \centering
  \includegraphics[width=\columnwidth]{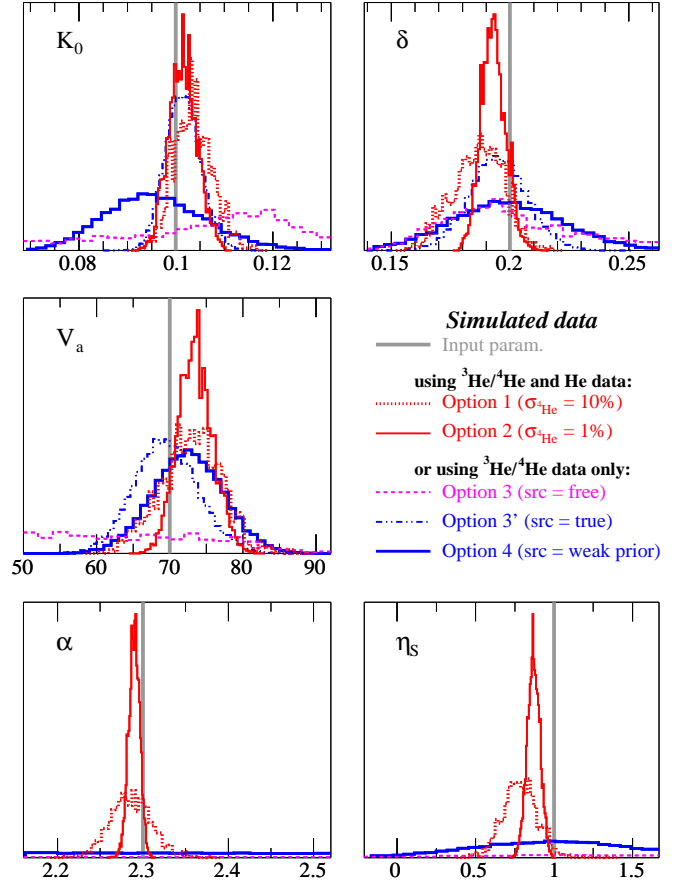}
  \caption{Marginalised posterior PDF for the transport and source parameters on the artificial data
  for Model~II
  (the values for the input model are shown as thick vertical grey lines in each panel). The colour code
  and style correspond to the five `options' described in Sect.~\ref{sec:strategy} (and used in
  Fig.~\ref{fig:ModelVc=0}).}
  \label{fig:PDF_Model_II}
\end{figure}

The MCMC analysis allows us to go further as it provides the PDF of the parameters,
from which the most-probable value and confidence intervals (CIs) are obtained. 
The results are gathered
in Table~\ref{tab:simu_param} and Fig.~\ref{fig:PDF_Model_II}. The various panels
of the latter represent the PDFs for transport and source parameters for each 
`option' for Model~II. (For concision the correlation plots are not shown.)
From these plots, some arguments are in favour of a simultaneous use of the secondary-to-primary
ratio and the primary flux (here, \hettohef{} and He), but not all.

\subsubsection{Advantages from a simultaneous analysis (ratio + flux)} 
A simultaneous analysis (\hettohef{} + He) gives more stringent constraints on the transport
parameters than only analysing the secondary-to-primary ratio (compare the PDFs for the red curves and blue
curves in Fig.~\ref{fig:PDF_Model_II} respectively, for $K_0$, $\delta$, and $V_c$). This partly comes from
the observed correlations between transport and source parameters\footnote{More
stringent constraints on the source parameters (from more precise data) leads
to more stringent constraints on the transport values: option~1 ($\sigma_{\rm He} \!=\! 10 \%$)
vs option~2 ($\sigma_{\rm He} \!=\! 1 \%$).}
(e.g., \citealt{2009A&A...497..991P}).
Option~2' and option~3' with fixed source parameters show that the better CIs on the transport parameters
come from the information contained in primary fluxes\footnote{Note that the lack
of constraints on the source parameters for option~4 confirms that the secondary-to-primary ratio
is only marginally sensitive to the source parameters (e.g., \citealt{2011A&A...526A.101P}).}.
The same conclusions hold true for Model~III ($V_c\neq0$), although with larger relative uncertainties
due to the two extra transport parameters of the model ($\eta_T$ and $V_c$).

\subsubsection{What if the wrong model is used?} As an illustration, we analyse
data simulated from model III ($V_c\neq 0$) with model II ($V_c=0$) and vice versa
(lower half of Table~\ref{tab:simu_param}). If we force $V_c=0$ (while $V_c^{\rm true}=18$~km~s$^{-1}$), the diffusion slope goes
to a low value $\delta\sim 0.2$ ($\delta^{\rm true}=0.7$), while the Alfv\'enic speed goes to
a high value $V_a\sim 100$~km~s$^{-1}$ ($V_a^{\rm true}=41$~km~s$^{-1}$). The 
larger $\chi^2_{\rm best}$/d.o.f. value with respect to the one obtained fitting the correct model
easily disfavours this model.
The second test (simulated with II, analysed with III) indicates whether allowing for more freedom in the analysis (two
additional free parameters $\eta_T$ and $V_c$) affects the recovery of the parameters. The values of
$\delta^{\rm true}=0.2$ and $V_a^{\rm true}=70$~km~s$^{-1}$ are recovered, while the others are systematically offset but
less than $3 \sigma$ away from their true value. 
In this simple example, adding extra parameters
is not an issue as the $\chi^2_{\rm best}$/d.o.f. still favours the minimal model.
However, with real data (see Sect.~\ref{sec:result_final} and the B/C analysis of 
\citealt{2010A&A...516A..66P,2010A&A...516A..67M}), in such a situation,
it is so far impossible to conclude whether the correct model is used, due
to the possible issue of multimodality and biases from systematics (see below).

\subsubsection{Drawbacks from a simultaneous analysis}
For Model~II but even more for Model~III (which has more free transport parameters),
a possible worry is the presence of multimodal PDF distributions, which more often
happens for the simultaneous analysis. An example of multimodality is
the analysis with Model II (i.e. $V_a=0$) of data simulated with Model~III,
which corresponds to a local minimum of the true Model III parameters.
This is of no consequence for the ideal case, but real data may
suffer from systematics errors and/or the inappropriate solar modulation model 
may be chosen. In that case, the true minimum can be displaced, or turned into
a local minimum (and vice versa).
Measurements over the last decades showed that primary fluxes are more prone
to systematics than secondary-to-primary ratios. Primary fluxes are also more
sensitive to solar modulation than ratios. For these reasons, the use of
secondary-to-primary data only (option~4) for the analysis,
although less performant to get stringent limits on the transport parameters,
is expected to be more reliable and robust.

\subsection{Recommended strategy to analyse real data\label{sec:strategy_final}}

The most robust approach to determine the transport parameters (and
their CIs) is to analyse the secondary-to-primary ratio using a loose but
physically-motivated prior on the source parameters (option~4). This has the advantage of
taking into account the correlations between the source and transport parameters. The
simultaneous analysis is mandatory to obtain the source parameters (option~1 or 2). It also brings
more information on the transport parameters, but the primary fluxes can bias their determination
if it suffers from systematics. We
recommend such an analysis to be performed in addition to the direct secondary-to-primary ratio
analysis, in order to get the following diagnosis: if the range of values for the
transport parameters from both analyses are
  \begin{itemize} 
    \item \emph{inconsistent}, it indicates that the values and CIs obtained for the sources
    parameters are biased or unreliable; 
    \item \emph{consistent}, the selected propagation model may be the correct one, and the
    source parameters are then the most probable ones for this model. However, the CIs on
    the transport parameters are very likely to be underestimated if the error bars on
    the ratio are much larger than the ones on the primary fluxes.
  \end{itemize}

Obviously, our analysis does not cover the range of all systematics when
dealing with real data. A more systematic analysis|e.g. covering a wider family of propagation
models, several solar modulation models, several sources of systematics in the data| goes
far beyond the scope of this paper. Note that some of these effects are likely to be
energy dependent, complicating even further the analysis. With the successful installation
of the AMS-02 detector on the ISS and its expected high-precision data, these issues
are bound to gain importance.

\section{Constraints from the quartet data\label{sec:res}}

We now apply the MCMC technique to the analysis of real data. We emphasise
that for the artificial ones, we assumed the \het{} to come solely from the \hef{} fragmentation, in order to
speed up the calculation. Based on our new compilation for the cross-section formulae (see
App.~\ref{app:quartet_xsec}), we take into account the contributions from $A>4$ CR parents, checking
which parents are relevant (\S\ref{sec:frac_contrib}). Having determined the heaviest parent to consider in the calculation,
we then move on to the result of the MCMC analysis (\S\ref{sec:real_results}), and those from
our best analysis (\S\ref{sec:result_final}).

\subsection{Fractional contributions\label{sec:frac_contrib}}

At first order, the contribution to the \deut{} and \het{} secondary production from $Z>4$ nuclei
is proportional to the source term ${\cal S}^j$ [see Eq.~(\ref{eq:CR})]. For a secondary contribution, the source term is proportional to the primary flux of the parents (which have been measured by many
experiments), and to the production cross-section. Normalised to the production from \hef{},
we have
\begin{equation}
   {\rm Rel}^{{{\rm P}\rightarrow\rm S}} \propto \frac{{\cal S}^{\rm P}}{{\cal S}^{\rm He}}\propto\frac{\Phi_{\rm P}}{\Phi_{\hef{}}} \cdot \gamma_{P}^{\rm S},
\end{equation}
where P is the CR projectile, S is the secondary fragment considered, and $\gamma_{\rm P}^{\rm S}$
[see Eq.~(\ref{eq:Nucp3})] is the production cross-section relative
to the production from \hef{}.
The fractional contribution $f^{\rm P\rightarrow\rm S}$ for each parent is defined to be
\begin{equation}
   f^{\rm P\rightarrow\rm S} = \frac{{\rm Rel}^{{{\rm P}\rightarrow\rm S}}}{\sum_{\rm P'=He\cdots Ni}{\rm Rel}^{{{\rm P'}\rightarrow\rm S}}}.
\end{equation}
As seen from Table~\ref{tab:contribs}, the most
important contributions from primary species heavier than He ($Z > 2$) are C and O, followed by Mg and Si and finally
Fe. The total contribution of these species amounts to $\sim35\%$ for \deut{} and $\sim11\%$ for \het{}, but
mixed species (such as N) or less abundant species also contribute to $\sim 5\%$.

\begin{table}[!tb]
\caption{Estimated fractional contribution of projectile $A>4$ to the \deut{} and \het{} fluxes. The columns are
respectively the name of the element, atomic number, ratio of measured fluxes, production cross-section
ratios for \deut{} and \het{}, and the estimated fractional contribution to \deut{} and \het{}.}
\label{tab:contribs}
\centering
\begin{tabular}{lrcccrr} \hline\hline
 P  &  $A_{\rm P}$   & $\frac{\Phi_{\rm P}}{\Phi_{\hef{}}}$ & $\gamma_{\rm P}^{\deut{}}$ & $\gamma_{\rm P}^{\het{}}$ & $f^{\rm P\rightarrow\deut{}}$ & $f^{\rm P\rightarrow\het{}}$ \vspace{-0.2cm}\\
        &          &                                      &                       &                        &              \%              &  \% \\
\hline
\multicolumn{4}{c}{\vspace{-0.15cm}} \\
He & 4  &       1.0       &$\cdots$&$\cdots$& 60.1 & 86.0   \vspace{0.1cm}\\
C  & 12 & $3.3\,10^{-2}$  &  5.5   &   2.3  & 7.5  & 3.6    \vspace{0.1cm}\\
N  & 14 & $7.4\,10^{-3}$  &  6.6   &   2.4  & 2.0  & 0.8    \vspace{0.1cm}\\
O  & 16 & $3.4\,10^{-2}$  &  7.8   &   2.6  & 10.7 & 4.1    \vspace{0.1cm}\\
F  & 19 & $5.2\,10^{-4}$  &  9.6   &   2.8  & 0.2  & 0.1    \vspace{0.1cm}\\
Ne & 22 & $5.1\,10^{-3}$  &  11.4  &   3.1  & 2.4  & 0.8    \vspace{0.1cm}\\
Na & 23 & $8.8\,10^{-4}$  &  12.1  &   3.2  & 0.4  & 0.1    \vspace{0.1cm}\\
Mg & 24 & $6.7\,10^{-3}$  &  12.7  &   3.3  & 3.4  & 1.0    \vspace{0.1cm}\\
Al & 26 & $1.1\,10^{-3}$  &  14.0  &   3.5  & 0.6  & 0.2    \vspace{0.1cm}\\
Si & 28 & $5.5\,10^{-3}$  &  15.3  &   3.7  & 3.4  & 1.0    \vspace{0.1cm}\\
P  & 31 & $1.8\,10^{-4}$  &  17.3  &   4.0  & 0.1  & $<0.1$ \vspace{0.1cm}\\
S  & 32 & $1.0\,10^{-3}$  &  17.9  &   4.1  & 0.7  & 0.2    \vspace{0.1cm}\\
Cl & 35 & $1.8\,10^{-4}$  &  20.0  &   4.5  & 0.1  & $<0.1$ \vspace{0.1cm}\\
Ar & 36 & $3.1\,10^{-4}$  &  20.6  &   4.6  & 0.3  & 0.1    \vspace{0.1cm}\\
K  & 39 & $2.1\,10^{-4}$  &  22.7  &   5.0  & 0.2  & $<0.1$ \vspace{0.1cm}\\
Ca & 40 & $6.1\,10^{-4}$  &  23.4  &   5.1  & 0.6  & 0.1    \vspace{0.1cm}\\
Sc & 45 & $1.0\,10^{-4}$  &  27.0  &   5.8  & 0.1  & $<0.1$ \vspace{0.1cm}\\
Ti & 48 & $3.4\,10^{-4}$  &  29.2  &   6.2  & 0.4  & 0.1    \vspace{0.1cm}\\
V  & 51 & $1.8\,10^{-4}$  &  31.4  &   6.6  & 0.2  & 0.1    \vspace{0.1cm}\\
Cr & 52 & $3.6\,10^{-4}$  &  32.1  &   6.8  & 0.5  & 0.1    \vspace{0.1cm}\\
Mn & 55 & $3.0\,10^{-4}$  &  34.3  &   7.2  & 0.4  & 0.1    \vspace{0.1cm}\\
Fe & 56 & $3.7\,10^{-3}$  &  35.1  &   7.4  & 5.3  & 1.3    \vspace{0.1cm}\\
Ni & 58 & $2.4\,10^{-4}$  &  36.6  &   7.7  & 0.4  & 0.1    \vspace{0.1cm}\\
\hline
\end{tabular}
\note{\tiny The ratio of measured fluxes is calculated at $\sim 10$ GeV/n using
PAMELA for He \citep{2011Sci...332...69A} and HEAO-3 \citep{1990A&A...233...96E} 
for $6\leq Z\leq30$.}
\end{table}
\begin{figure}[!t]
\centering 
  \includegraphics[width=\columnwidth]{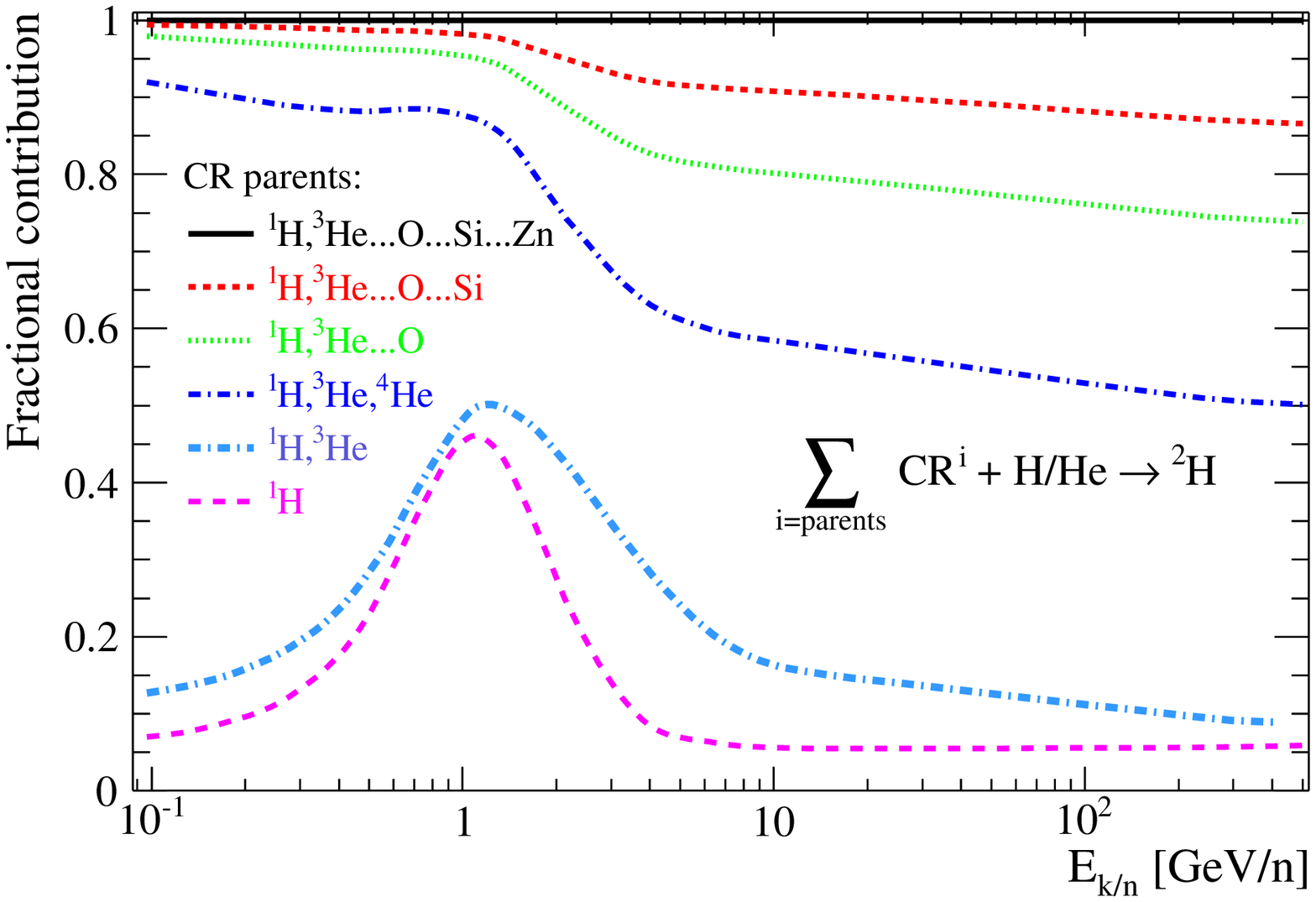}  
  \includegraphics[width=\columnwidth]{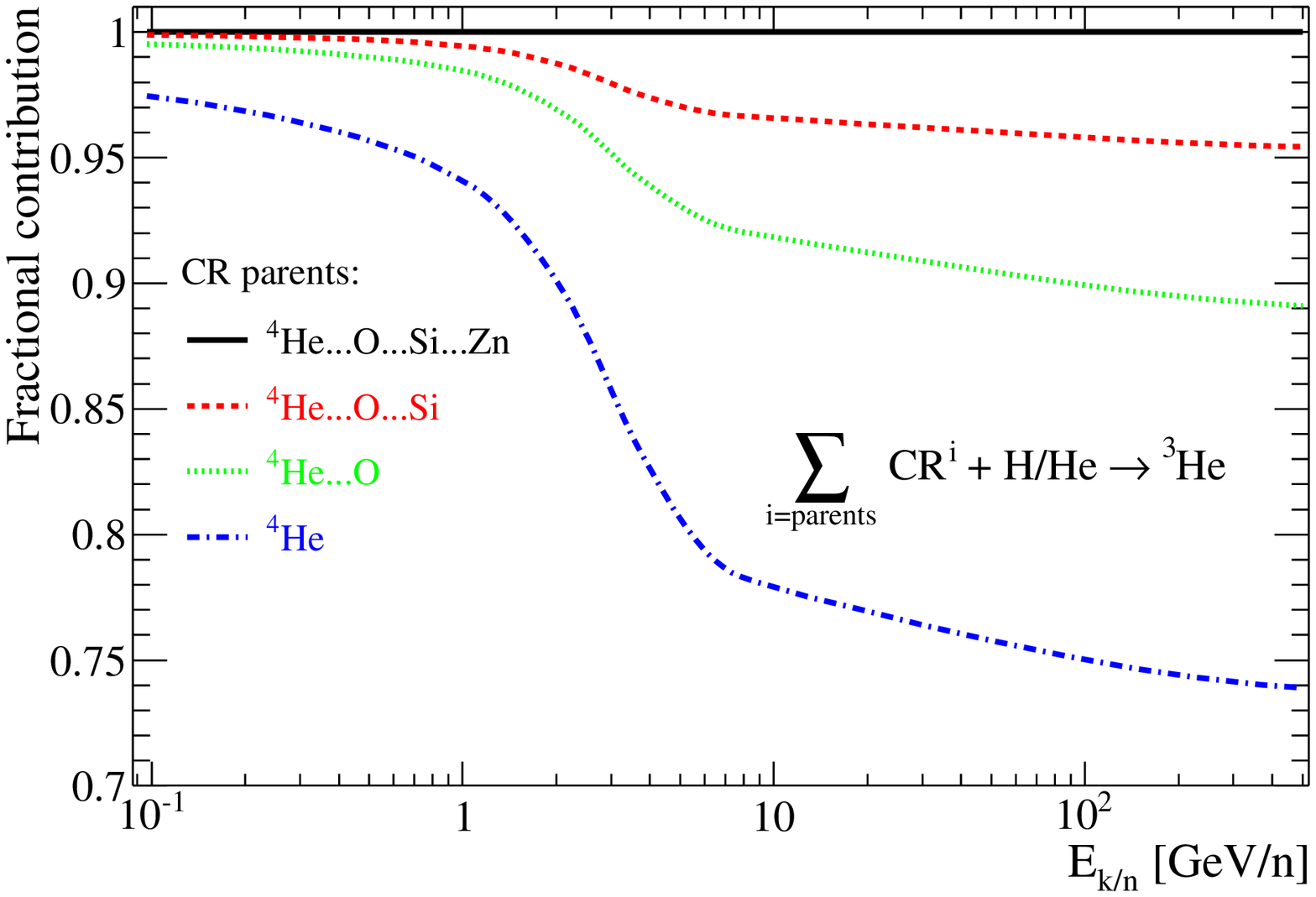} 
  \includegraphics[width=\columnwidth]{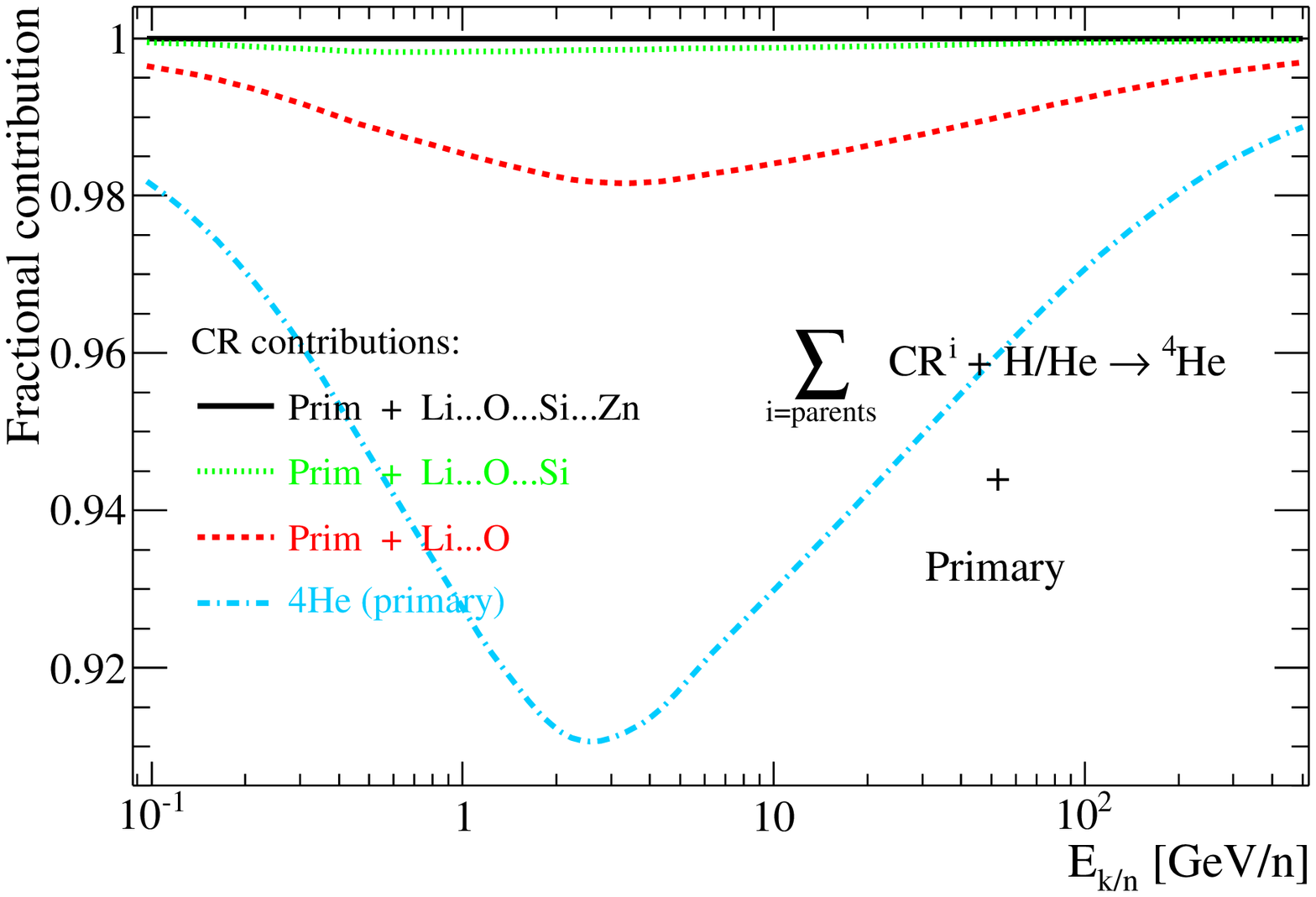}  
  \caption{Fractional contributions to the propagated \deut{} fluxes (top panel), 
  \het{} fluxes (middle panel) and \hef{} fluxes (bottom panel) as a function of
  $E_{k/n}$ from $A>4$ CR parents. For \hef{}, the primary contribution is also considered.} 
  \label{fig:FractionalContribution3He} 
\end{figure} 

A proper calculation of these fractional contributions involves the full solution of the propagation
equation, taking into account energy gains and losses, total inelastic reactions, and convection. Based
on the propagation parameters found to fit best the current data (see next section), we show in
Fig.~\ref{fig:FractionalContribution3He} the fractional contribution of $A>4$ nuclei as a function of
energy for the full calculation. It confirms the previous figures, but with a residual energy dependence
(itself depending on the species) hitting a plateau above $\sim 100$~GeV/n. The difference 
can be mostly attributed to a preferential destruction of heavier nuclei at low
energy\footnote{Indeed, the primary-to-primary ratios are not constant. The heavier the nucleus,
the larger its destruction cross-section, the more the propagated flux is affected/decreased
at low energy, the longer it takes to reach a plateau of maximal contribution at high energy.
The observed trend is consistent with the primary-to-primary ratios shown in Fig.~14
of \citet{2011A&A...526A.101P}.}. 
Note that for \deut{} production, the coalescence of two protons (long-dashed pink curve)
contributes up to 40\% of the total at $\sim 1$ GeV/n energy (peak of the cross-section,
see Fig.~\ref{fig:sigprod2H_other}).
Depending on the precision reached for the data, it is important to include the CNO
contribution \citep[e.g.][]{1969ApJ...155..587R,1973ICRC....5.3086J,1986ApJ...311..425B}, but also all
contributions up to Ni.

Finally, the fragmentation of CNO can also affect the \proton{} and \hef{} primary fluxes.
The peak of contribution occurs at
GeV/n as secondary fluxes drop faster than primary fluxes with energy. 
Fig.~\ref{fig:FractionalContribution3He} shows this contribution to be $\lesssim 10\%$ for \hef{}.
With the high precision measurement from PAMELA and the even better measurements awaited from AMS-02,
this will need to be further looked into in the future.

\subsection{MCMC analysis: test of several data combinations\label{sec:real_results}}

\begin{table*}[!t]
\caption{MCMC analysis of Model III ($V_c\neq 0$) with $L=4$~kpc: most-probable values and relative uncertainties
(corresponding to the 68\% CI) for the analysis of various combinations of \hettohef{}, \het{}, and PAMELA He data
(+ 3 combinations involving the \deut{} isotope). The last column gives the $\chi^2$/d.o.f. for the best-fit model found (which usually differs from the most-probable one).}
\label{tab:BestFitParameters3He}\centering
\begin{tabular}{lcccccccc} \hline\hline
Data  & $K_{0} \times 10^{2}$      & $\delta$ &    $V_{c}$     &     $V_a$      & $\alpha$  & $\eta_{S}$  &   $\chi^2_{\rm best}$/d.o.f. \\
      & \!\! (kpc$^2$\,Myr$^{-1})$ &    -     & (km\,s$^{-1}$) & (km\,s$^{-1}$) &     -     &     -       &      -     \\
\hline
& \multicolumn{4}{c}{\vspace{-0.25cm}} \\
\hettohef{} \  + He                & $0.50^{+ 10\%}_{-12\%}$   & $0.79^{+6.4\%}_{-5.2\%}$  & $17.3^{+1.7\%}_{-2.5\%}$   & $39.5^{+6.1\%}_{-6.6\%}$  & $2.26^{+1.8\%}_{-2.7\%}$  & $ 0.07^{+59\%}_{-74\%}$   & 4.8 \vspace{0.1cm}\\
\hettohef{} \ + He$^\ddagger$            & $1.10^{+27\%}_{-27\%}$    & $0.54^{+5.6\%}_{-7.4\%}$  & $28.0^{+3.6\%}_{-3.6\%}$   & $54.0^{+5.6\%}_{-9.3\%}$  & $2.49^{+0.4\%}_{-0.8\%}$  & $-1.97^{+9.6\%}_{-1.0\%}$ & 2.1 \vspace{0.1cm}\\
\hettohef{}                        & $0.50^{+84\%}_{-50\%}$    & $0.67^{+6.0\%}_{-15\%}$   & $27.4^{+3.3\%}_{-4.7\%}$   & $41.0^{+34\%}_{-19\%}$    & $[1.8,2.5]$                        & $[-2,+2]$       & 2.9 \vspace{0.1cm}\\
& \multicolumn{4}{c}{\vspace{-0.1cm}} \\
\hettohef{} \ + \het{}             & $ 1.20^{+ 58\%}_{-8.3\%}$ & $ 0.56^{+11.\%}_{-8.9\%}$  & $24.2^{+4.5\%}_{-2.9\%}$  & $68.1^{+6.8\%}_{-16.\%}$  & $2.23^{+2.7\%}_{-3.6\%}$  & $-0.48^{+33\%}_{-52\%}$   & 1.8 \vspace{0.1cm}\\
\hettohef{} \ + \het{}$^\P$        & $ 2.82^{+ 35\%}_{-50\%}$  & $ 0.39^{+26.\%}_{-5.1\%}$  & $24.4^{+6.1\%}_{-8.6\%}$  & $85.0^{+7.1\%}_{-22.\%}$  & $2.17^{+2.3\%}_{-4.6\%}$  & $-0.82^{+18\%}_{-34\%}$   & 1.9 \vspace{0.1cm}\\
\hettohef{} \  + \het{} + He       & $ 0.76^{+ 10\%}_{-9.2\%}$ & $ 0.79^{+3.8\%}_{-2.5\%}$  & $19.6^{+1.4\%}_{-2.8\%}$  & $48.1^{+4.0\%}_{-4.8\%}$  & $2.22^{+0.9\%}_{-1.8\%}$  & $0.02^{+171\%}_{-159\%}$  & 3.3 \vspace{0.1cm}\\
\hettohef{} \ + \het{}$^\P$  + He  & $ 0.81^{+ 10\%}_{-8.6\%}$ & $ 0.77^{+5.2\%}_{-2.6\%}$  & $19.2^{+1.6\%}_{-3.1\%}$  & $48.2^{+5.0\%}_{-3.7\%}$  & $2.21^{+1.4\%}_{-1.4\%}$  & $0.04^{+ 71\%}_{-63\%}$   & 3.4 \vspace{0.1cm}\\
& \multicolumn{4}{c}{\vspace{-0.1cm}} \\
\deuttohef{} \ + \deut{}           & $   15^{+ 13\%}_{-33\%}$  & $ 0.03^{+100\%}_{-21\%}$    & $ 3.0^{+500\%}_{-67\%}$  & $33.7^{+52\%}_{-28\%}$    & $ 2.38^{+4.6\%}_{-3.8\%}$ & $-1.15^{+17\%}_{- 35\%}$  & 6.8 \vspace{0.1cm}\\
\deuttohef{} \ + \deut{} + He      & $ 0.53^{+ 17\%}_{-15\%}$  & $ 0.61^{+8.2\%}_{-6.6\%}$   & $14.6^{+3.4\%}_{-2.7\%}$ & $26.3^{+8.0\%}_{-11\%}$   & $ 2.47^{+1.2\%}_{-2.4\%}$ & $ 0.47^{+ 13\%}_{-21\%}$  & 6.4 \vspace{0.1cm}\\

\deuttohef{} \  + \deut{} + He$^\ddagger$  & $6.5^{+55\%}_{-45\%}$ & $0.039^{+125\%}_{-125\%}$ & $26.1^{+8\%}_{-27\%}$ & $24.1^{+39\%}_{-46\%}$ & $2.74^{+11\%}_{-11\%}$ & $0.38^{+47\%}_{-39\%}$ & 4.9 \vspace{0.1cm} \\

\hline
\end{tabular}
{\footnotesize \\ 
  $^\P$ Including AMS-01 data from the recently published analysis of \citet{2011ApJ...736..105A}.\\
  $^\ddagger$ Excluding PAMELA He data points below 5 GeV/n and above 183 GeV/n.}
\note{\tiny An interval in square brackets corresponds to the prior used for the analysis (the posterior PDF obtained is close to the prior).}
\end{table*}

Given the accuracy of current data (see Fig.~\ref{fig:envelopes}), we must take into account
the contribution from all parent nuclei at least up to $^{30}$Si. In the rest of the analysis,
we use PAMELA data for He \citep{2011Sci...332...69A}, as they overcome
all others in the $\sim$~GeV$-$TeV range in terms of precision. Before giving our final results,
and to complement Sect.~\ref{sec:strategy}, we discuss the appropriate choice of data
to consider here, in order to get the best balance between robustness and reliability for the \deut{}
and \het{}-related analyses.

\subsubsection{Simulated {\em vs.} real data}

We start by comparing the results obtained with the simulated and the actual data set. To
avoid lengthy comparisons of numbers, we limit ourselves to Model~III (where we also fix
$\eta_T$ to its default value, i.e. 1). The obvious difference with the simulated data is
that we no longer have access to the true source parameters (automatically excluding
options 2' and 4 discussed in Sect.~\ref{sec:strategy}). 
For the simultaneous analysis using He PAMELA data|the precision of which is $\sim1\%$|, we recover
similar values and CIs for the parameters (compare option~2 in Table \ref{tab:simu_param}
and the first three lines of Table~\ref{tab:BestFitParameters3He}). The second row
of Table~\ref{tab:BestFitParameters3He} is based on a subset of He data: high energy data points
are discarded because they show departure from a single power-law \citep{2010ApJ...714L..89A,2011Sci...332...69A},
whereas low-energy data points are discarded because of their sensitivity to solar modulation, which is presumably
too crudely described by the Force-Field approximation used here. The $\chi^2_{\rm best}$/d.o.f. value
(first row)
shows that the model has difficulty to perfectly match the high precision PAMELA He data over the whole energy range. The analysis of
\hettohef{} ratio using a prior on the source parameters (option~4 in Table \ref{tab:simu_param}
and third line of Table~\ref{tab:BestFitParameters3He}) gives larger CIs for the transport parameters.
The consistency between the results of the latter analysis (third line) and that based on the partial He data (second line),
and their discrepancy with the results of the analysis based on the full He data set (first line)
confirms our suspicion that high precision measurements for primary fluxes can bias the
transport parameters determination.

\subsubsection{Adding the secondary \het{} flux in the analysis}
Replacing He by \het{} in the simultaneous analysis ($4^{\rm th}$ and $5^{\rm th}$ line)
further affects the determination of the transport
parameters. This is not surprising since \het{} data are not all consistent with one
another (see Fig.~\ref{fig:envelopes}). The bias is stronger when taking into account
the recently published AMS-01 data \citep{2011ApJ...736..105A}. If both \het{} and He
are taken into account\footnote{The simultaneous analysis of \hettohef{} + \het{} + He has
not been tested in the simulated data since it would have amounted to a double-counting of
the \het{} data (appearing in the three quantities). However, real data involve different
experiments for the various quantities (PAMELA for He and other experiments for \het{}), and
independent measurements are used.}, the much better accuracy of the He
PAMELA data with respect to the \het{} data amounts to a smaller weight of the
latter in the analysis.

\subsubsection{\deuttohef{} {\em vs} \hettohef{}} 

We repeat partially the analysis for \deut{} in the last 3 rows of
Table~\ref{tab:simu_param}. The data are so inconsistent with one another for \deuttohef{} (see
Fig.~\ref{fig:envelopes}) that we are forced to use at least the \deut{} flux (whose
data points are also markedly inconsistent with one another). Even so, the results
are not reliable. PAMELA and AMS-02 have the capability
to improve greatly the situation, but in the meantime, we are forced to include He as well in the analysis
(next-to-last row in the table). The transport parameter values
from the \deuttohef{}+\deut{}+He analysis are grossly consistent with those from the \hettohef{}+\het{}+He analysis,
but are likely to suffer from similar biases (see the previous paragraph).
Reducing the energy range of He data is not even possible for the \deut{} analysis (last line in the table),
as the results obtained are not reliable.

\begin{table*}[!t]
\caption{Most-probable values and CIs for Models II and III for our `best' analysis (Sect.~\ref{sec:result_final})
of the quartet and B/C data ($L=4$~kpc).}
\label{tab:BestFitParametersConclusion}\centering
\begin{tabular}{lcccccccc} \hline\hline
Data                                    & $K_{0} \times 10^{2}$      &       $\delta$                &    $V_{c}$                &     $V_a$            &        $\alpha$           &       $\eta_{S}$         & $\chi^2_{\rm best}$/d.o.f. \\
                                 & $\!\!\!$(kpc$^2$ Myr$^{-1})\!\!\!$&            -                  &    (km\,s$^{-1}$)         &    (km s$^{-1}$)     &           -               &            -             &          -         \\
\hline 
\multicolumn{8}{c}{\vspace{-0.2cm}} \\
\multicolumn{8}{c}{\em Model II\vspace{0.05cm}} \\
\hettohef{}                             & $15.0^{+0.5}_{-0.5}$       & $0.29^{+0.02}_{-0.03}$        &              -            & $116^{+11}_{-7}$&     $[1.8,2.5]$           & $[-2,+2]$               & 3.3   \\[0.1cm]
\hettohef{} + \het{} + He$^\ddagger$    & $7.0^{+0.2}_{-0.3}$        & $0.31^{+0.03}_{-0.02}$        &              -            & $74^{+4}_{-3}$  & $2.36^{+0.03}_{-0.01}$    & $0.43^{+0.08}_{-0.05}$  & 4.6   \\[0.1cm]
\deuttohef{} + \deut{} + He             & $14.8^{+0.5}_{-0.5}$    & $0.08^{+0.03}_{-0.03}$     &              -            & $44^{+5}_{-8}$  & $2.66^{+0.03}_{-0.03}$    & $0.70^{+0.05}_{-0.03}$  & 5.6   \\[0.1cm]
B/C     ~~~~~~~~~[Putze et al., 2010]   & $9.7^{+0.3}_{-0.2}$        & $0.234^{+0.006}_{-0.005}$      &              -            & $73^{+2}_{-2}$        & $\alpha+\delta=2.65$      &          $-1$             & 4.7   \\[0.1cm]
B/C   ~~~~~~~~~~~~~~~~~~~[this paper]   & $6.2^{+0.4}_{-0.3}$        & $0.35^{+0.01}_{-0.01}$        &              -            & $80^{+2}_{-2}$        &     $[1.8,2.5]$           &     $[-2,+2]$           & 1.5   \\[0.1cm]
B/C + C	(all)      ~~~~~~[this paper]   & $6.5^{+0.1}_{-0.1}$        & $0.314^{+0.006}_{-0.006}$     &              -            & $57^{+2}_{-1}$        & $2.340^{+0.005}_{-0.008}$ & $0.96^{+0.04}_{-0.04}$  & 13.9  \\[0.1cm]
B/C + C	(HEAO)           [this paper]   & $6.3^{+0.1}_{-0.1}$        & $0.353^{+0.004}_{-0.004}$     &              -            & $78^{+1}_{-2}$        & $2.250^{+0.015}_{-0.006}$ & $1.48^{+0.08}_{-0.12}$  & 2.8   \\[0.1cm]
\hline
\multicolumn{8}{c}{\vspace{-0.2cm}} \\
\multicolumn{8}{c}{\em Model III\vspace{0.05cm}} \\
\hettohef{}                             & $0.5^{+0.4}_{-0.3}$     & $0.67^{+0.04}_{-0.10}$        & $27.4^{+0.9}_{-1.3}$      & $41^{+14 }_{-8}$      &$[1.8,2.5]$                & $[-2,+2]$               & 2.9   \\[0.1cm]
\hettohef{} + \het{} + He$^\ddagger$    & $1.0^{+0.3}_{-0.2}$     & $0.64^{+0.03}_{-0.05}$        & $23.5^{+0.5}_{-1.0}$      & $54^{+5}_{-3}$        & $2.37^{+0.03}_{-0.03}$    & $0.03^{+0.04}_{-0.04}$  & 1.6   \\[0.1cm]
\deuttohef{} + \deut{} + He             & 3.2$^{+0.8}_{-0.1}$        & $0.50^{+0.08}_{-0.10}$         & $27.1^{+2.5}_{-1.8}$      & $72^{+9}_{-16}$ & $2.41^{+0.12}_{-0.08}$    & $0.39^{+0.07}_{-0.06}$  & 5.5   \\[0.1cm]
B/C     ~~~~~~~~~[Putze et al., 2010]   & $0.46^{+0.08}_{-0.06}$     & $0.86^{+0.04}_{-0.04}$        & $18.9^{+0.3}_{-0.4}$      & $38^{+2}_{-2}$        & $\alpha+\delta=2.65$      &          $-1$             & 1.5   \\[0.1cm]
B/C   ~~~~~~~~~~~~~~~~~~~[this paper]   & $0.46^{+0.18}_{-0.10}$     & $0.82^{+0.08}_{-0.05}$        & $18.3^{+0.2}_{-0.3}$      & $40^{+5}_{-4}$        & $[1.8,2.5]$               & $[-2,+2]$               & 0.9   \\[0.1cm]
B/C + C	(all)      ~~~~~~[this paper]   & $0.57^{+0.05}_{-0.03}$     & $0.80^{+0.02}_{-0.01}$        &  $17.4^{+0.2}_{-0.3}$     & $36^{+1}_{-1}$        & $2.260^{+0.007}_{-0.009}$ & $0.24^{+0.03}_{-0.05}$  & 5.2   \\[0.1cm]
B/C + C	(HEAO)           [this paper]   & $0.33^{+0.06}_{-0.10}$     & $0.93^{+0.05}_{-0.07}$        &  $18.2^{+0.3}_{-0.2}$     & $35^{+4}_{-2}$        & $2.312^{+0.019}_{-0.008}$ & $1.9^{+0.1}_{-0.2}$     & 2.0   \\[0.1cm]
\hline
\end{tabular}
{\footnotesize \\
    $^\ddagger$ Excluding PAMELA He point below 5 GeV/n and above 183 GeV/n.
}
\note{\tiny An interval in square brackets corresponds to the prior used for the analysis (the posterior PDF obtained is close to the prior).}
\note{\tiny The B/C results are based on IMP7-8~\citep{1987ApJS...64..269G}, Voyager~1\&2 \citep{1999ICRC....3...41L},
ACE-CRIS \citep{2009ApJ...698.1666G}, HEA0-3 \citep{1990A&A...233...96E}, Spacelab~\citep{1991ApJ...374..356M}, AMS-01~\citep{2011ApJ...736..105A},
and CREAM \citep{2008APh....30..133A}, shown to be the most compatible data for a B/C analysis \citep{2009A&A...497..991P}.}
\end{table*}

\subsection{MCMC analysis: `best' results\label{sec:result_final}}

\begin{figure*}[!t] 
\centering
\includegraphics[width=\columnwidth]{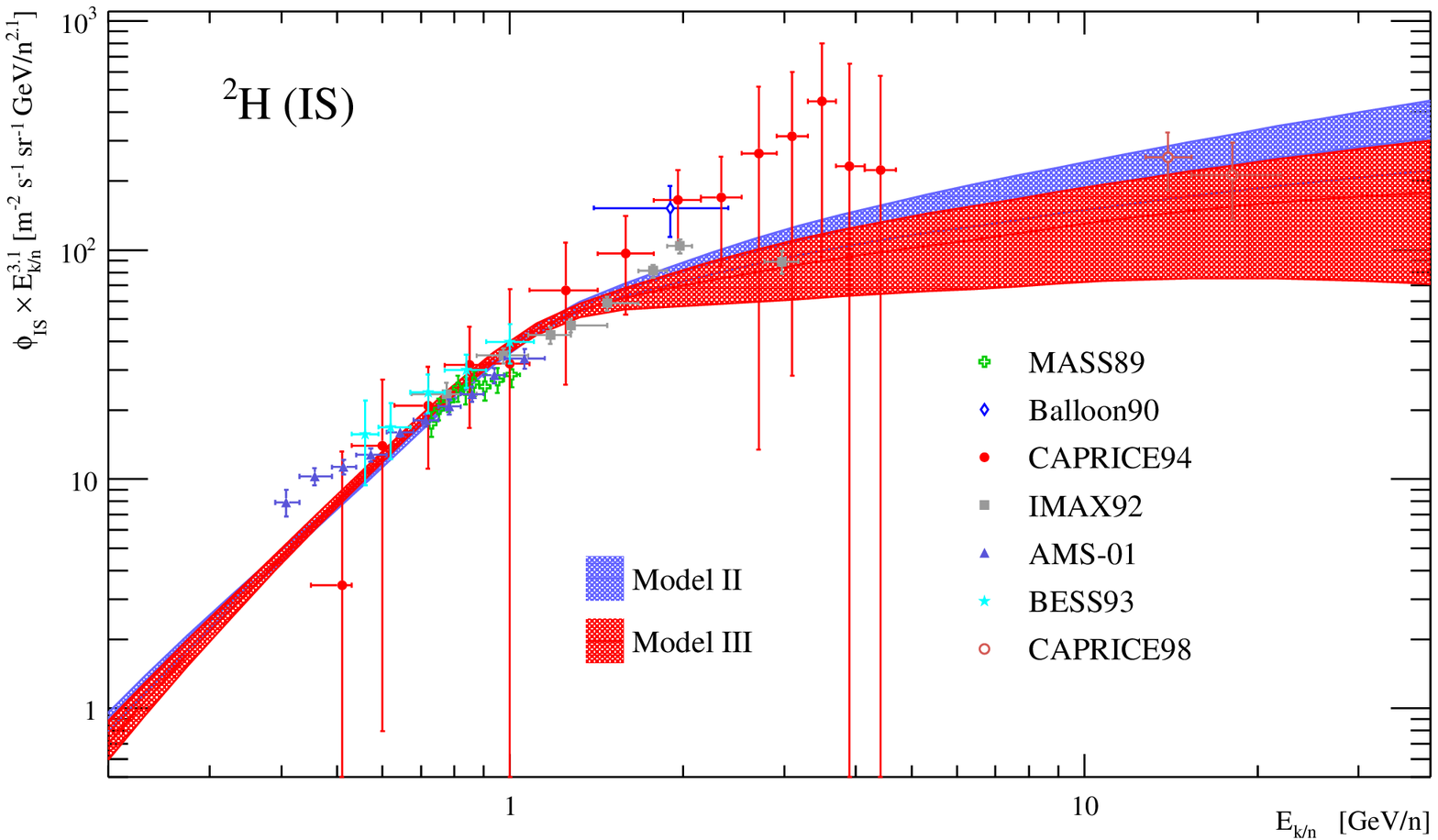}
\includegraphics[width=\columnwidth]{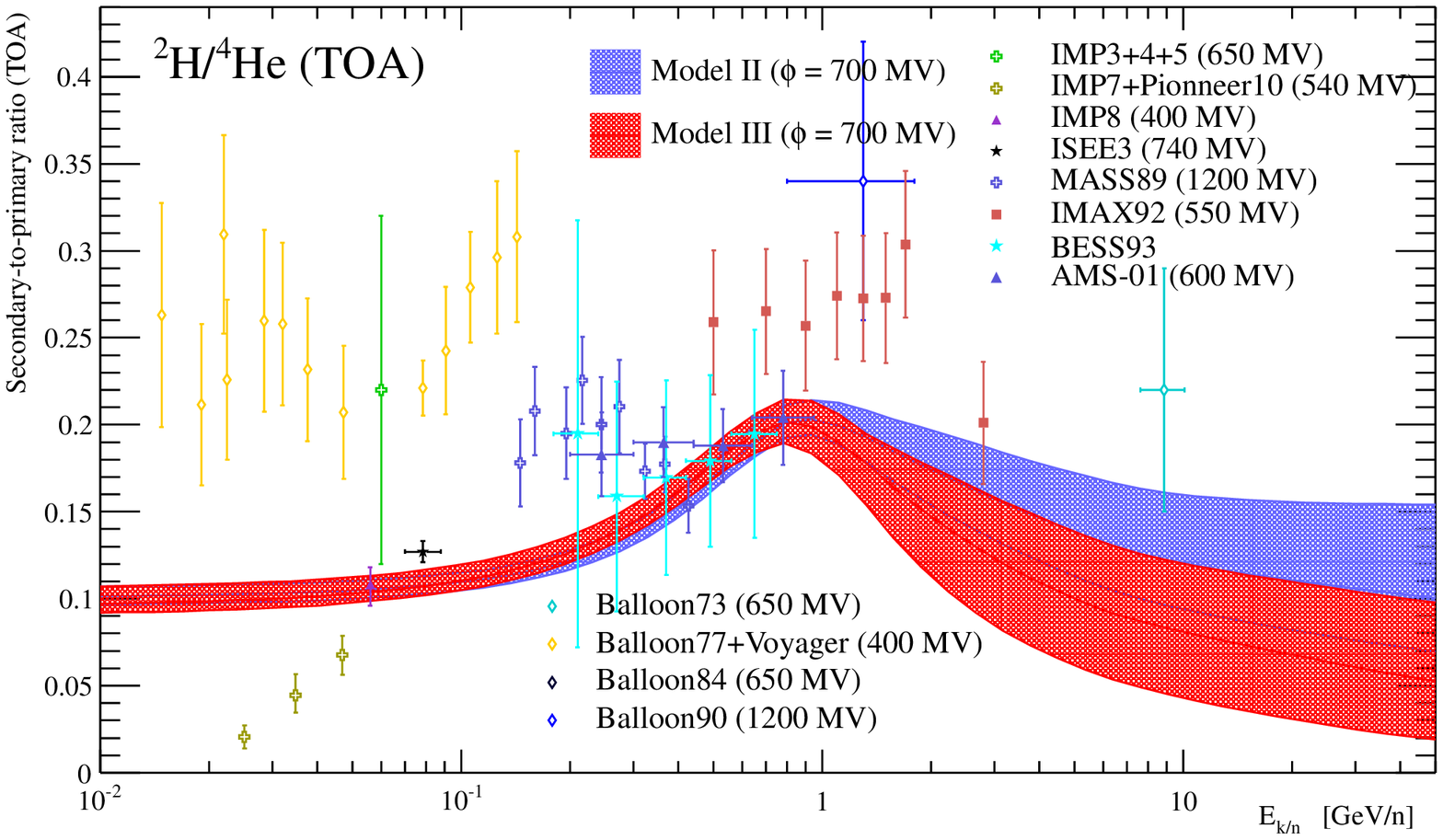}
\includegraphics[width=\columnwidth]{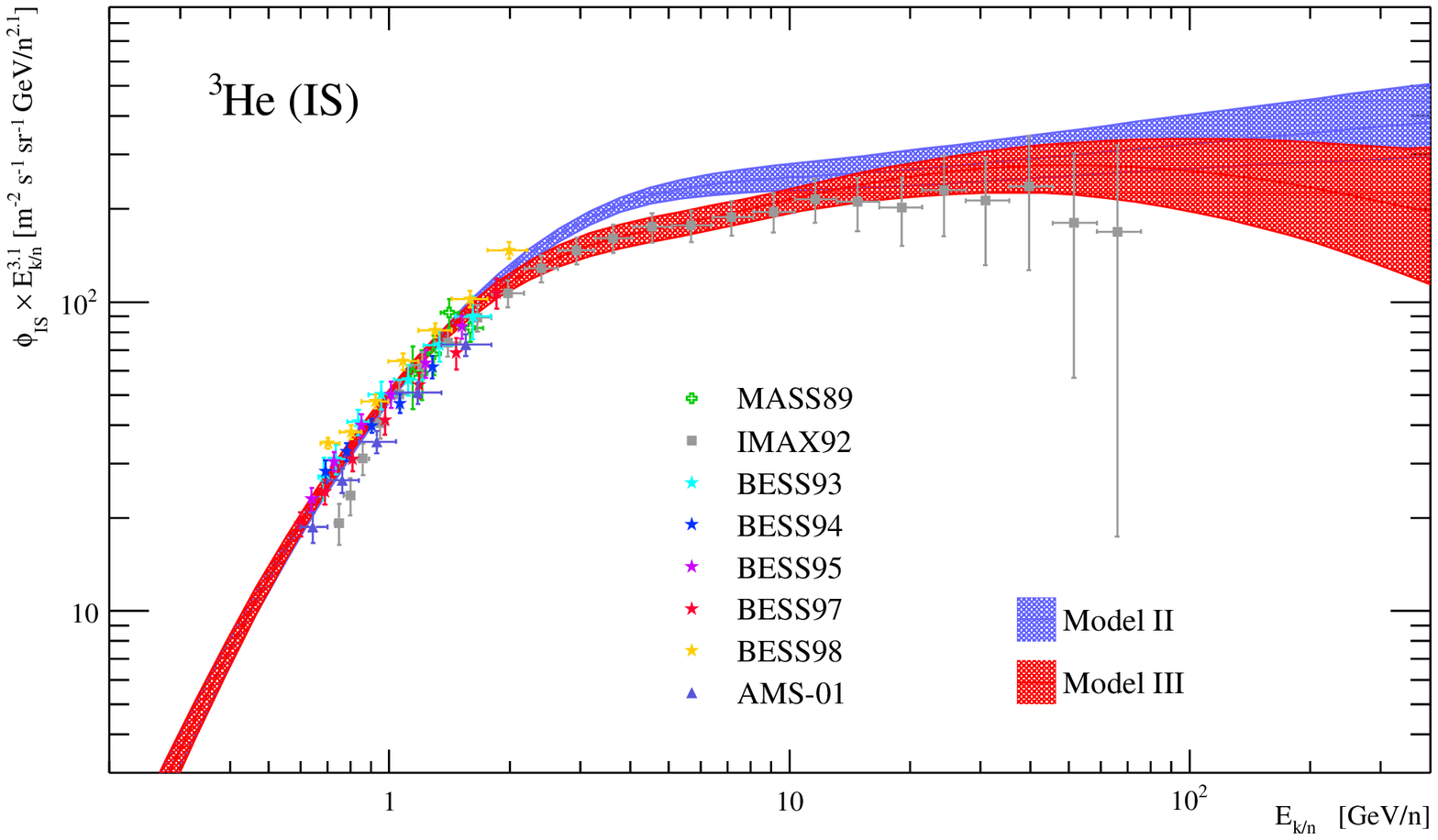}
\includegraphics[width=\columnwidth]{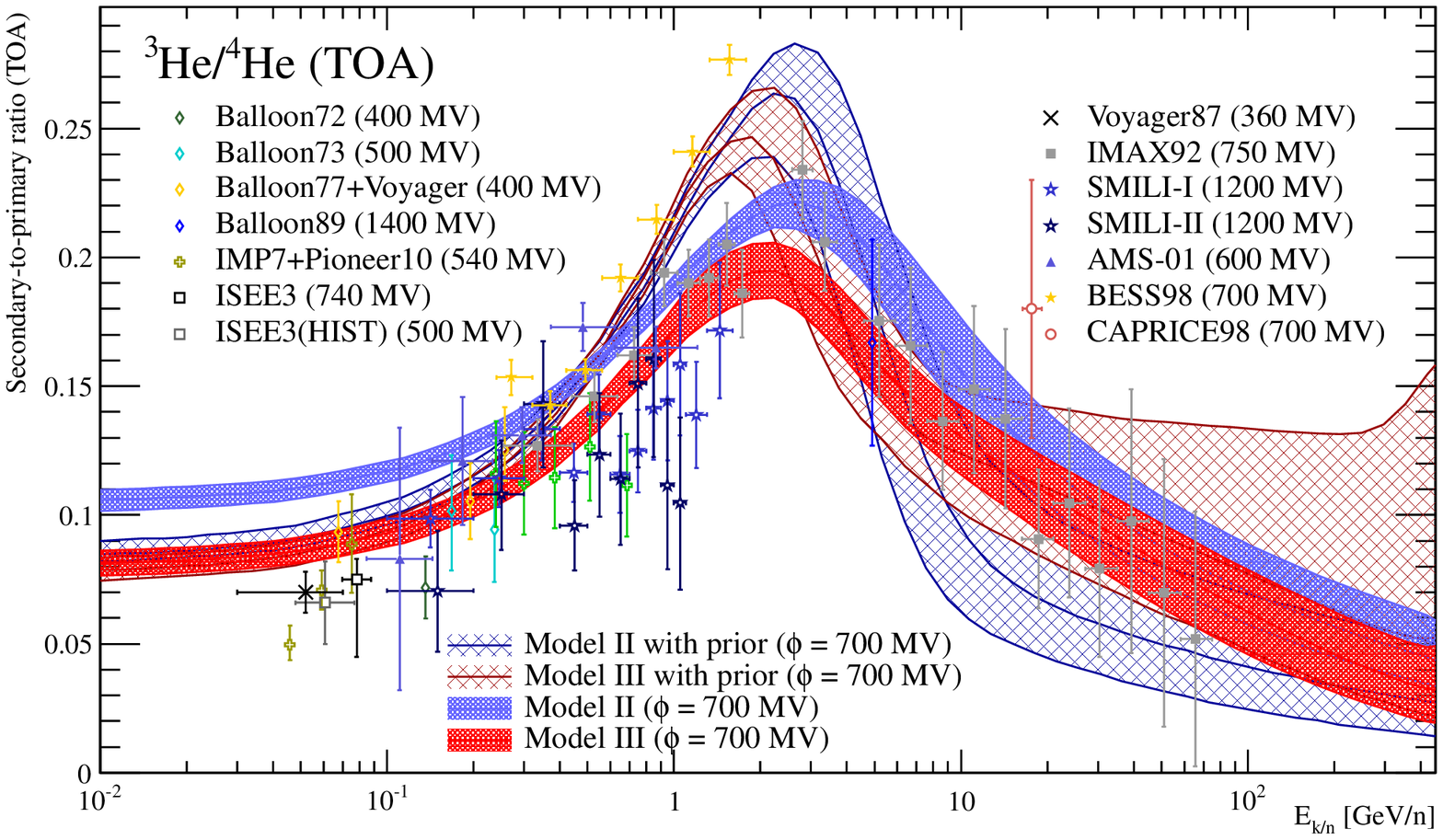}
\caption{Left panels: demodulated interstellar \deut{} (top) and \het{} (bottom)
envelopes at 95\% CIs times $E_{k/n}^{3.1}$. Right panels: top-of-atmosphere secondary-to-primary
ratio \deuttohef{} (top) and \hettohef{} (bottom)
ratios. The full envelopes correspond to the `best' simultaneous
analysis (secondary-to-primary ratio + primary flux) for model II (blue) and III (red).
The hatched envelopes correspond to the \hettohef{} analysis (prior on source parameters).
See Table~\ref{tab:refs_quartet} for references and the corresponding
demodulation (for IS) and modulation (TOA) level $\phi$.}
\label{fig:envelopes}
\end{figure*} 

Taking into account specificities of the actual data (previous section), our `best' analysis
is based on the most relevant combinations of data for \deut{} and \het{}:
\begin{itemize}
  \item the \hettohef{} analysis (with a prior for the source parameters) gives robust and
   conservative results for the transport parameters. The result from the
   \hettohef{}+\het{}+He$^\ddagger$ analysis is more sensitive to biases,
   but using an energy sub-range for He data is expected to limit them.
  \item Due to the paucity of \deut{} data, the \deuttohef{}+\deut{}+He (full energy-range
  for He) analysis is the only reliable option, although it probably suffers from biases.
  \end{itemize}
The corresponding most-probable values and CIs are gathered in Table~\ref{tab:BestFitParametersConclusion},
and the corresponding envelopes for \deuttohef{}, \deut{}, \hettohef{}, and \hettohef{} are given
in Fig.~\ref{fig:envelopes}. We also re-analyse the B/C ratio according to our `best-analysis' 
scheme (B/C alone with a prior for the source parameters or B/C + C). The results are reported in 
Table~\ref{tab:BestFitParametersConclusion}, where the results obtained
in \citet{2010A&A...516A..66P} for fixed source parameters are also reproduced:
we note that the new strategy gives results in better agreement
with those of the quartet analysis (e.g., the transport parameters $\delta$ and $K_0$
are shifted by more than 30\% for Model~II), further demonstrating its usefulness.

\subsubsection{Universality of the transport parameters}
If we focus on the transport parameters, we note that combinations involving the \deuttohef{},
\hettohef{}, or B/C ratio give broadly consistent  transport parameter values, be it for Model~II
or Model~III\footnote{The most significant difference is for the \deuttohef{}+\deut{}+He analysis,
which is inconsistent in both models and clearly unreliable for Model~II ($\delta\sim 0$).
For model III, B/C and \hettohef{}-related constraints are roughly in the same region but are located
at several $\sigma$ from each others (they are consistent with one another for Model II).}.
Regardless of the actual propagation model, we
conclude that these results hint at the universality of CR transport for all species.
Another important result is that the constraints set by the quartet data on the transport
parameters are competitive with those set by the B/C ratio, so that the quartet data should be a prime target
for AMS-02.

\subsubsection{Model~II ($\delta\sim0.3$) or Model~III ($\delta\sim0.7$)?} 
According to Sect.~\ref{sec:strategy}, comparing the results of the secondary-to-primary
ratio analysis with those of the combined analysis (ratio + primary flux) gives an indication
of their robustness. Table~\ref{tab:BestFitParametersConclusion} show that the results for
the diffusion slope $\delta$ is very robust, regardless of the model considered.
A more detailed comparison shows that for \het{}-related constraints, the transport parameter values
for Model~II are inconsistent with one another at the $3 \sigma$ level, whereas  the 68\% CIs overlap with
one another (but for $V_c$) for Model~III, hence slightly favouring the latter ($\delta~\sim 0.7$). 
The comparison of the $\chi^2_{\rm best}$/d.o.f. values also tends to favour model III. 
Hence, although the value $\delta\sim 0.7$ seems favoured, we cannot exclude yet pure reacceleration model
($V_c=0$) with $\delta\sim 0.3$. Moreover, as shown in \citet{2010A&A...516A..67M}, many ingredients
of the propagation models can lead to a systematic scatter of the transport parameters larger than
the width of their CIs. Data at higher energy for any secondary-to-primary ratio are mandatory to
conclude on this issue.

\subsubsection{Source spectrum}
The present analysis is more general than that used in \citet{2011A&A...526A.101P},
where the transport parameters were fixed. Although it is not the main focus
of this paper, we remark that the values of the source slope $\alpha$
from the B/C + C analysis are consistent with those found in \citet{2011A&A...526A.101P},
strengthening the case of a universal source slope $\alpha$ at the $\sim 5\%$ level.
For the quartet values, $\alpha_{\rm He}$ is broadly consistent with \citeauthor{2011A&A...526A.101P}'s analysis
(based on AMS-01, BESS98 and BESS-TeV data for He). However, the results for the source parameters
depend on the choice of data sets and energy-range considered. This indicates that for $\lesssim1\%$
accuracy data, either the model for the source is inappropriate, or the solar modulation model is faulty, or some systematics
exist in the measurements. The AMS-02 data will help to clarify this question.

\section{Conclusion\label{sec:conclusion}}

We have revisited the constraints set on the transport (and also the source) parameters by the quartet data,
i.e. \proton{}, \deut{}, \het{}, and \hef{} fluxes, but also the secondary-to-primary ratios \deuttohef{}
and \hettohef{}. This extends and complements a series of studies
\citep{2009A&A...497..991P,2010A&A...516A..66P,2011A&A...526A.101P} carried out with the USINE propagation
code and an MCMC algorithm. The three main ingredients on which the analysis rests are:

\begin{itemize}
  \item A minute compilation of the existing quartet data and survey of the literature, showing that the
most recent/precise data (AMS-01, BESS93$\rightarrow$98, CAPRICE98, IMAX92, and SMILI-II) have not been
considered before this analysis.
  \item We have done a systematic survey of the literature for the cross-sections involved in the
production/survival of the quartet nuclei. This has lead us to propose new empirical production cross-sections
of \deut{}, \trit{}, and \het{}, valid above a few tens of MeV/n, for any projectile on p and He (we also
updated inelastic cross-sections).
  \item We have made an extensive use of artificial data sets to assess the reliability of the  derived CIs of
  the GCR transport and source parameters for various combinations of data/parameters analyses.
\end{itemize}

In broad agreement with previous studies, \citep[e.g.][]{1969ApJ...155..587R,1986ApJ...311..425B}, we find
that the fragmentation of CNO contributes significantly to the \deut{} flux ($\sim 30\%$) above a few
GeV/n energies (\hef{} fragmentation is the dominant channel for the \deut{} and \het{} fluxes).
Nevertheless, we provide a much finer picture, showing in particular that heavy nuclei ($8<Z\le30$)
contribute up to 10\% for \het{} (20\% for \deut{}) at high energy. We also provide an estimate of the
secondary fraction to the \hef{} flux. By definition, the secondary contribution has a steeper spectrum
than the primary one and therefore becomes quickly negligible at high energy. This secondary contribution
peaks at a few GeV/n, and it amounts to $\sim 10\%$ of the total flux ($\sim 7\%$ up to O fragmentation,
$\sim 2\%$ from elements heavier than O), which is already a sizeable amount given the $\sim 1\%$
precision reached by the PAMELA data~\citep{2011Sci...332...69A}. For \proton{}, the knowledge of the
multiplicity of neutron and proton produced by the interaction of all elements on the ISM is required to
calculate precisely its secondary content.

Simulated data have allowed us to check several critical behaviours. Firstly, the He flux is obviously useful
(and required) to constrain the source parameters, but it has also been found to bring significant information
on the transport parameters: fitting a secondary-to-primary ratio plus a primary flux brings more
constraints than just fitting the ratio (even when source parameters are fixed). Secondly, we have checked 
that a model with more free parameters (than the ones used to simulate the data) is able to recover the
correct values. However, our analysis has also strongly hinted at the fact that adding the primary flux He
biases the determination of the transport parameters if systematics (which are usually more important in
primary fluxes than in ratios) are present, and/or if the wrong model is used. For this reason, when dealing
with measurements, we recommend to always compare the result from the {\em secondary-to-primary ratio + primary flux}
analysis to that of the {\em secondary-to-primary ratio} using a loose but physically motivated prior on the source
parameters.

The analysis of real data has shown that quartet data slightly favours a model with large
$\delta\sim 0.7$ (with $V_c\sim 20$~km~s$^{-1}$ and $V_a\sim 40$~km~s$^{-1}$), but that
a model with small $\delta\sim 0.2$ (with $V_c\sim 0$ and  $V_a\sim 80$~km~s$^{-1}$) cannot be completely
ruled out. Better quality data, and especially data at higher energy are required to go further.
The conclusions are similar and the range of transport parameters found are consistent with
those obtained from the B/C analysis
\citep{2001ApJ...547..264J,2010A&A...516A..66P,2010A&A...516A..67M}\footnote{Note that in this paper, we
did not attempt to combine the results of different secondary-to-primary ratios (\deuttohef{},
\hettohef{}, B/C, sub-Fe/Fe, $\bar{p}/p$). This is left for a future study, for which a Bayesian evidence
could be used to better address (in a Bayesian framework) the crucial issue of model selection.}.
This strongly hints at the the universality of the GCR transport for any all nuclei. Furthermore,
we have shown that the analysis of the light isotopes (and the already very good precision on He)
is as constraining as the B/C analysis (similar range of CIs).

The several difficulties which have been pointed out in this analysis could be alleviated by virtue
of using better data. However, it is more likely that the interpretation of future high-precision data will
require the development of refined models for the source spectra and/or transport and/or solar
modulation. For instance, the Force-Field approximation for solar modulation is already too crude to
minutely match the PAMELA He data. The forthcoming AMS-02 data at an even better accuracy will definitively pose interesting
new challenges.

\begin{acknowledgements}
D.~M. would like to thank A.~V. Blinov for kindly providing copies of
several of his articles. A.~P. is grateful for financial support from the Swedish Research Council (VR) through the Oskar Klein
Centre.

\end{acknowledgements}

\appendix

\section{Cosmic-ray data\label{app:quartet}}

Deuterons and \het{} fluxes are very sensitive to the modulation level, whereas
ratios are less affected. The exact value for the solar modulation level
$\phi$ is uncertain. For instance, the values given in the
seminal papers can differ greatly from those estimated by \citet{2004ApJ...612..262C}
(in order to match the electron and positron fluxes of various experiments, see
their Table~1), or from those reconstructed from the Neutron monitors 
\citep{2002SoPh..207..389U}. This difference may be attributed to the fact that
the latter analysis correctly solves the Fokker-Plank equation of GCR transport
in the Heliosphere, whereas most papers rely on the widely used force-field
approximation that is known to fail for strong modulation level $\phi\gtrsim 1000$~MV
\citep[e.g.,][]{2002SoPh..207..389U}. In this analysis, we do not attempt
to go beyond this force-field approximation, as speed is of essence for
our MCMC analysis. We rely mostly on the force-field effective modulation
parameter $\phi$ necessary to reproduce the data (as quoted in the seminal paper),
but these values are slightly adjusted in order to give overlapping fluxes
when all the data are demodulated and plotted together. Given the uncertainty
on the data, the large uncertainty on $\phi$, and the fact that most-probable
region of parameter space is constrained by the \deuttohef{} and \hettohef{}
ratio (rather than the best fit to the \deut{} and \het{} fluxes), we feel that
it is a safe procedure till high precision data from PAMELA of AMS-02 are
available.

The demodulated interstellar (IS) fluxes for \deut{} and \het{} are
shown in the  left panels of Fig.~\ref{fig:envelopes},
whereas the yet modulated top-of-atmosphere (TOA) ratios for \deuttohef{}
and \hettohef{} are shown in its right panels. The references for the data
are given in Table~\ref{tab:refs_quartet}.
\begin{table*}[!t]
\caption{References for the quartet data.}
\label{tab:refs_quartet}
\centering
\begin{tabular}{lrcrll} \hline\hline
  Exp.       & \#data& Year &$\!\!\!\!\phi$ (MV)& Ref.  & Comment \\\hline
& \multicolumn{1}{c}{} \vspace{-0.30cm} \\
      \multicolumn{6}{c}{--- \deut{} ---} \vspace{0.05cm}\\  
MASS89       &   9   & 1989 &   1200    & \citet{1991ApJ...380..230W}  &\vspace{0.025cm}\\ 
Voyager87    &   7   & 1987 &  360      & \citet{1994ApJ...432..656S}  & Voyager is at 23 AU \vspace{0.025cm}\\ 
Balloon90    &   1   & 1990 &  1200     & \citet{1995ICRC....2..598B}  &\vspace{0.025cm}\\ 
Voyager94    &  4    & 1994 & 150       & \citet{1995ApJ...451L..33S}  &Voyager is at 56 AU \vspace{0.025cm}\\ 
CAPRICE94    &  14   & 1994 & 600       & \citet{1999ApJ...518..457B}  &Subtraction of H to \proton{} from Tab~3.\vspace{0.025cm}\\ 
IMAX92       &   8   & 1992 & 550       & \citet{2000AIPC..528..425D}  &\vspace{0.025cm}\\ 
AMS-01       &   10  & 1998 & 600       & \citet{2002PhR...366..331A}  &\vspace{0.025cm}\\ 
BESS93       &   5   & 1993 & 700       & \citet{2002ApJ...564..244W}  &\vspace{0.025cm}\\ 
CAPRICE98    &   5   & 1998 & 700       & \citet{2004ApJ...615..259P}  &\vspace{0.15cm}\\ 
      \multicolumn{6}{c}{--- \het{} ---} \vspace{0.05cm}\\  
MASS89       &   5   & 1989 &   1200    & \citet{1991ApJ...380..230W}  &\vspace{0.025cm}\\ 
Voyager87    &   7   & 1987 &  360      & \citet{1994ApJ...432..656S}  &Voyager is at 23 AU \vspace{0.025cm}\\ 
Voyager94    &  4    & 1994 & 150       & \citet{1995ApJ...451L..33S}  &Voyager is at 56 AU \vspace{0.025cm}\\ 
IMAX92       &  24   & 1992 & 750       & \citet{2000ApJ...533..281M}  &\vspace{0.025cm}\\ 
BESS93       &   7   & 1993 & 700       & \citet{2002ApJ...564..244W}  &\vspace{0.025cm}\\ 
BESS94       &   5   & 1994 & 630       & \citet{2003ICRC....4.1805M}  &Taken from their Fig.~2\vspace{0.025cm}\\ 
BESS95       &  6    & 1995 & 550       & \citet{2003ICRC....4.1805M}  &Taken from their Fig.~2\vspace{0.025cm}\\ 
BESS97       &  7    & 1997 & 491       & \citet{2003ICRC....4.1805M}  &Taken from their Fig.~2\vspace{0.025cm}\\ 
BESS98       &  7    & 1998 & 700       & \citet{2003ICRC....4.1805M}  &Taken from their Fig.~2\vspace{0.025cm}\\ 
AMS-01       &  5    & 1998 & 600       & \citet{2011ApJ...736..105A}  &\vspace{0.15cm}\\ 
      \multicolumn{6}{c}{--- \deuttohef{} ---}\vspace{0.05cm} \\ 
IMP3+4+5       &   1   & 65+67+69  &   650     & \citet{1971ApJ...166..221H}  & \vspace{0.025cm}\\
Balloon73      &   1   &    1973   &   650     & \citet{1973ICRC....1..126A}  & \vspace{0.025cm}\\
IMP7+Pioneer10$\!\!\!\!$&  3    &   72+73   &   540     & \citet{1975ApJ...202..815T}  & \vspace{0.025cm}\\
Balloon77+Voyager$\!\!\!\!$&  14   &   1977    &   400     & \citet{1983ApJ...275..391W}  & \vspace{0.025cm}\\
 IMP8          &  1    &   1977    &   400     & \citet{1985ApJ...294..455B}  & \vspace{0.025cm}\\
 ISEE3         &  1    &  78-84    &   740     & \citet{1986ApJ...303..816K}  & \vspace{0.025cm}\\
Balloon84      &  1    &   1974    &   650     & \citet{1988AA...189...51D}  & (discarded in the analysis)\vspace{0.025cm}\\
MASS89         &  9    &   1989    &  1200     & \citet{1991ApJ...380..230W}  & \vspace{0.025cm}\\
Balloon90      &  1    &   1990    &  1200     & \citet{1995ICRC....2..598B}  & \vspace{0.025cm}\\
IMAX92         &  8    &   1992    & 550       & \citet{2000AIPC..528..425D}  &\vspace{0.025cm}\\ 
AMS-01         &  4    &   1998    & 600       & \citet{2011ApJ...736..105A}  &\vspace{0.15cm}\\ 
      \multicolumn{6}{c}{--- \hettohef{} ---}\vspace{0.05cm} \\ 
Balloon72      &  2    &  1972     &   400     & \citet{1975ICRC....1..312W}  & Re-analysed by \citet{1987ApJ...312..178W}\vspace{0.025cm}\\ 
IMP7+Pioneer10$\!\!\!\!$&  2    &   72+73   &   540     & \citet{1975ApJ...202..815T}  & \vspace{0.025cm}\\
Balloon73      &  2    &  1973     &   500     & \citet{1978ApJ...221.1110L}  &\vspace{0.025cm}\\ 
Balloon77+Voyager$\!\!\!\!$&  3    &   1977    &   400     & \citet{1983ApJ...275..391W}  & Re-analysed by \citet{1987ApJ...312..178W}\vspace{0.025cm}\\
Balloon81      &   1   &   1981    &   440     & \citet{1985ApJ...291..207J}  & (discarded, see \citealt{1987ApJ...312..178W})\vspace{0.025cm}\\ 
 ISEE3         &   2   &  78-84    &   740     & \citet{1986ApJ...303..816K}  & \vspace{0.025cm}\\
 ISEE3(HIST)   &   1   &    78     &   500     & \citet{1986ApJ...311..979M}  & \vspace{0.025cm}\\
MASS89         &   5   &   1989    &  1200     & \citet{1991ApJ...380..230W}  & \vspace{0.025cm}\\
SMILI-I        &  12   &   1989    &  1200     & \citet{1993ApJ...413..268B}  &\vspace{0.025cm}\\ 
Voyager87      &   1   &   1987    &  360      & \citet{1994ApJ...432..656S}  &Voyager is at 23 AU \vspace{0.025cm}\\ 
Balloon89      &   1   &   1989    &  1400     & \citet{1995PhRvD..52.6219H}  &\vspace{0.025cm}\\ 
IMAX92         &  21   &   1992    & 750       & \citet{2000ApJ...533..281M}  &\vspace{0.025cm}\\ 
SMILI-II       &  10   &   1991    &  1200     & \citet{2000ApJ...534..757A}  &\vspace{0.025cm}\\ 
BESS98         &   7   &   1998    & 700       & \citet{2003ICRC....4.1805M}  & \vspace{0.025cm}\\ 
CAPRICE98      &   1   &   1998    & 700       & \citet{2003ICRC....4.1809M}  &\vspace{0.025cm}\\ 
AMS-01         &   5   &   1998    & 600       & \citet{2011ApJ...736..105A}  & Supersedes \citet{Xiong:2003hp} data \vspace{0.075cm}\\ 
\hline
\end{tabular}
\end{table*}

\section{Cross-sections\label{app:quartet_xsec}}

This appendix summarises the production and destruction cross-sections
employed for the quartet nuclei in this paper.

\subsection{Elastic and inelastic cross-sections}

All reaction cross-sections are taken from the parametrisations of
\citet{1999STIN...0004259T}, but for the pH reaction cross-section. The latter
is evaluated from  $\sigma^{\rm inel}_{\rm pp}=\sigma^{\rm tot}_{\rm pp} - \sigma^{\rm
el}_{\rm pp}$, where the total and elastic cross-sections are fitted to the data
compiled in the PDG\footnote{\tt http://pdg.lbl.gov/}.  Note also that for
$\hef{} + \hef{}$, we had to renormalise \citet{1999STIN...0004259T} formulae by a
factor 0.9 to match the low-energy data.

\begin{figure}[!t] 
\centering
\includegraphics[width=\columnwidth]{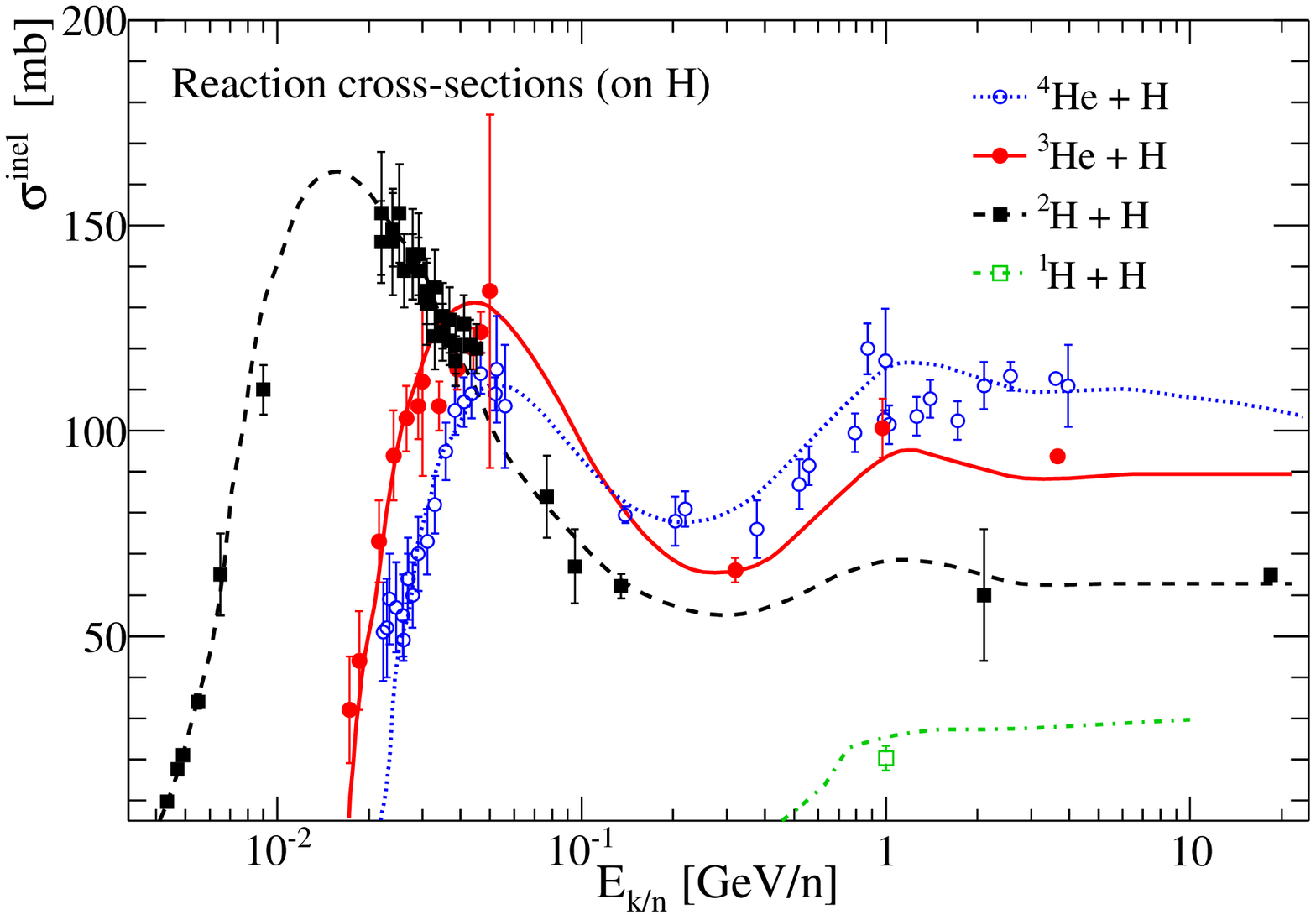}
\includegraphics[width=\columnwidth]{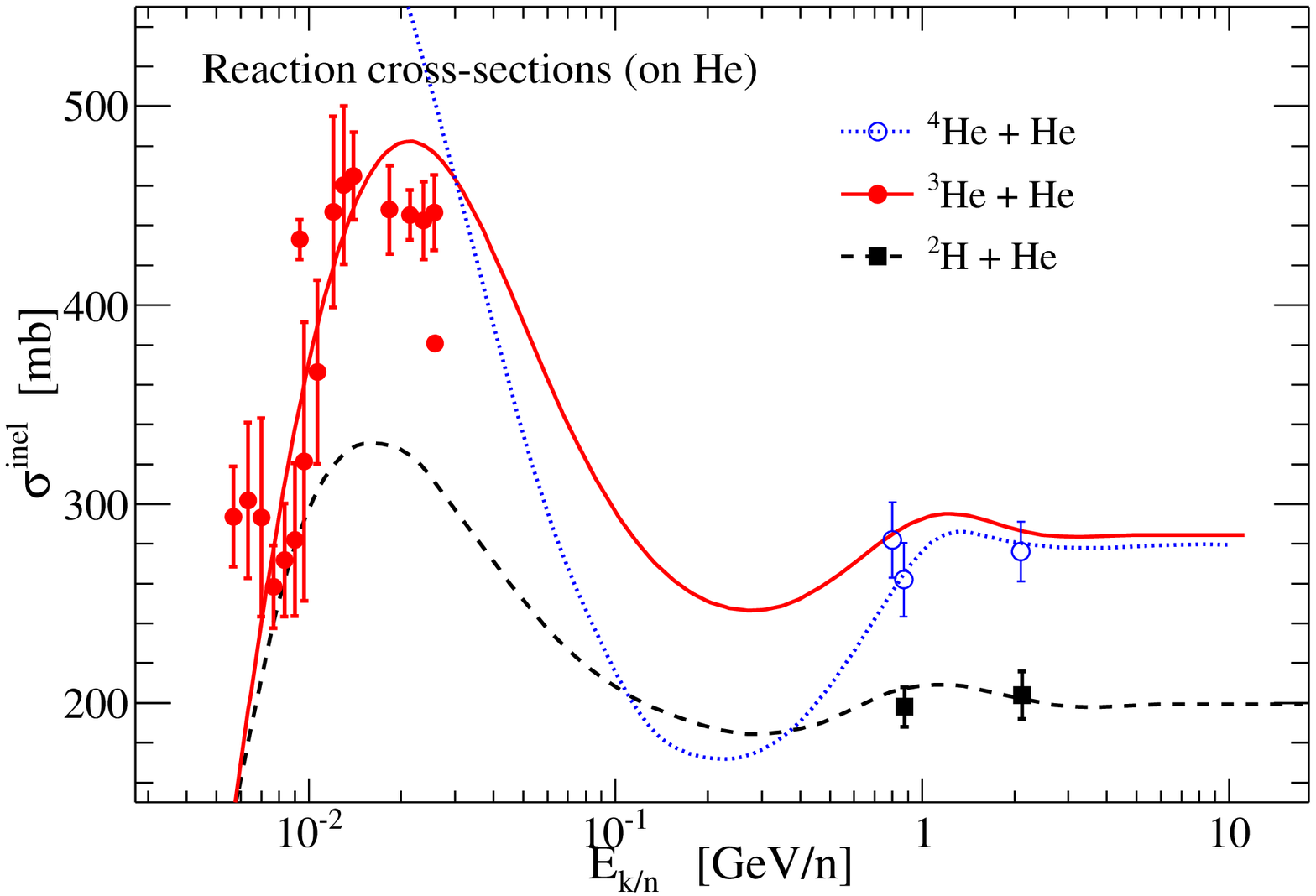}
\caption{Total inelastic (reaction) cross-section for the quartet isotopes.
The lines show our parametrisation (see text), the symbols are data.
{\bf Top panel:} reaction on H with data from \citet{Cairns1964,Hayakawa1964,Igo1967,1967PhRvL..18.1200P,1969ApJ...158..711G,Nicholls1972,Carlson1973,1976PhRvC..13..451S,Ableev:1977ua,1977PhLB...70..155K,1978PhRvC..18.2273J,Velichko1982,Blinov:1984zk,Blinov:1984kx,1993PAN....56..536A,Glagolev:1993cw,1997AdSpR..19..755W}.
{\bf Bottom panel:} reaction on He with data from \citet{1977PhRvC..16...18K,1978PhRvC..18.2273J,Tanihata:1985zq}.
}
\label{fig:Xsec_inel}
\end{figure} 

Our parametrisations (lines) and the data (symbols) are shown in Fig.~\ref{fig:Xsec_inel} for reaction on H and He.
Note that we rely on \citet{1997lrc..reptQ....T} for any other inelastic reaction.

\subsection{Light nuclei production: Nuc + p}

The light nuclei \het{} and \deut{} are spallative products of cosmic rays interacting with
the interstellar medium (ISM). The total secondary flux is obtained from the combination of
production cross-sections and measured primary fluxes. In principle, all nuclei must be
considered, but the ISM and GCRs are mostly composed of \proton{} and \hef{}, making the
reactions involving these species dominant. For heavier species, their decreasing number is
balanced by their higher cross-section. In several studies
\citep[e.g.][]{1969ApJ...155..587R,1973ICRC....5.3086J,1986ApJ...311..425B}, it was found
that the CNO$_{\rm CR}+$ H$_{\rm ISM}$ reactions contribute to $\sim~30\%$ of the \deut{} flux
above GeV/n energies. The reverse reaction  H$_{\rm CR}+$CNO$_{\rm ISM}$ mostly produces
fragments at lower energies, making them irrelevant for CR studies in the regime $\gtrsim
100$ MeV/n\footnote{Solar modulation also ensures that only species created at energies
$\gtrsim$ GeV/n matter.}. Note that \trit{} is also produced in these reactions, but it
decays in \het{} with a life time ($12.2$ years) short with respect to the
propagation time. All tritium production is thus assimilated to \het{} production,
but the cross-sections for this fragment are provided as well below.

The energy of the fragments roughly follows a Gaussian distribution \citep[e.g.][]{Cucinotta}.
Its impact on the secondary flux was inspected for the B/C analysis by \citet{1995ApJ...451..275T},
where an effect $\lesssim 10\%$ was found, compared to the straight-ahead approximation,
in which the kinetic energy per nucleon of the fragment equals that of the projectile.
The precision sought for the cross-sections is driven by the level of precision
attained by the CR data to analyse. Given the large errors on the existing 
data, the straight-ahead approximation is enough for this analysis.
However, future high-precision data (e.g. from the AMS-02 experiment) will probably
require a refined description.

\subsubsection{\hef{} + p $\rightarrow$ \deut{}, \trit{}, and \het{}}

\begin{figure}[!t]
\begin{center}
\includegraphics[width=\columnwidth]{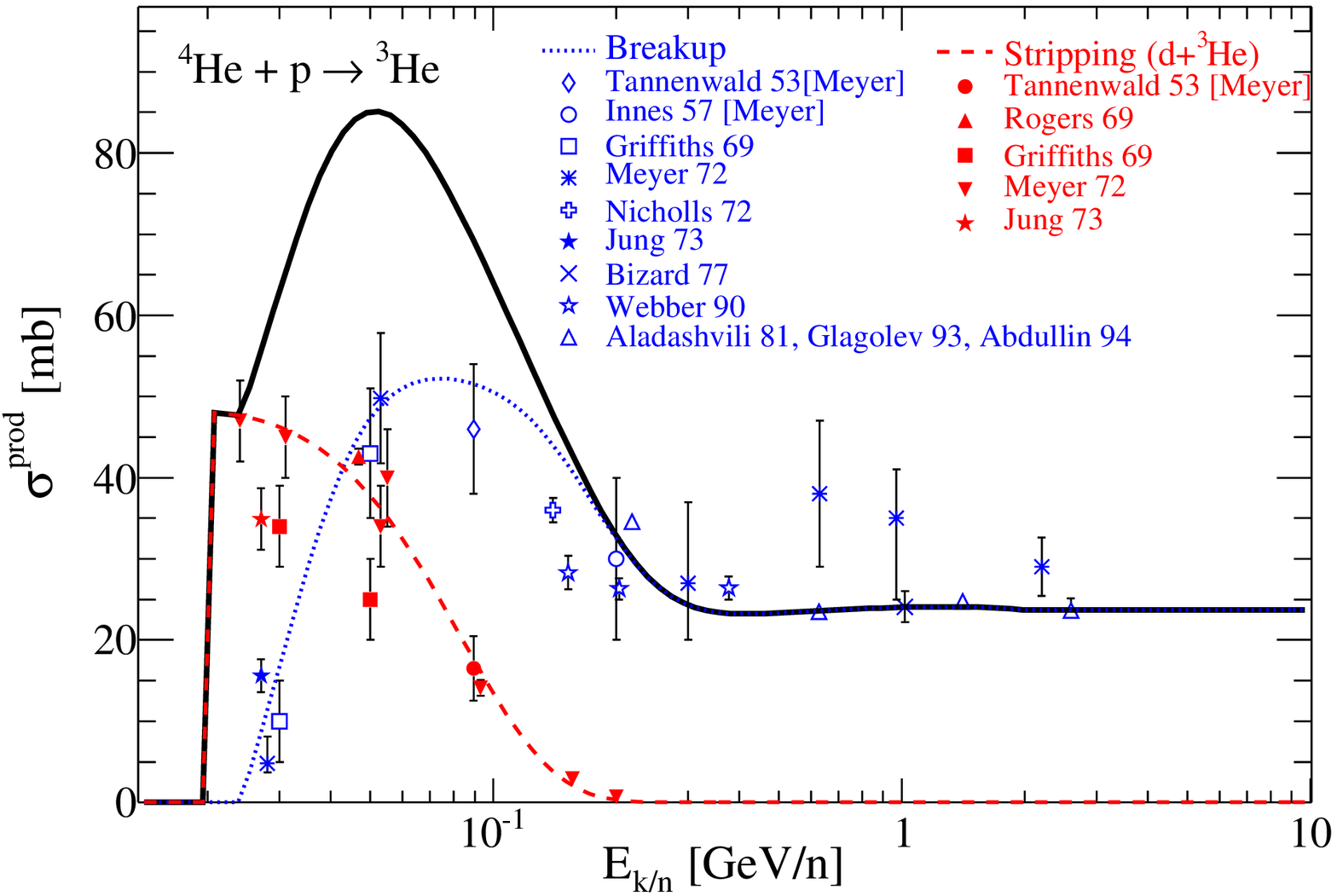}
\includegraphics[width=\columnwidth]{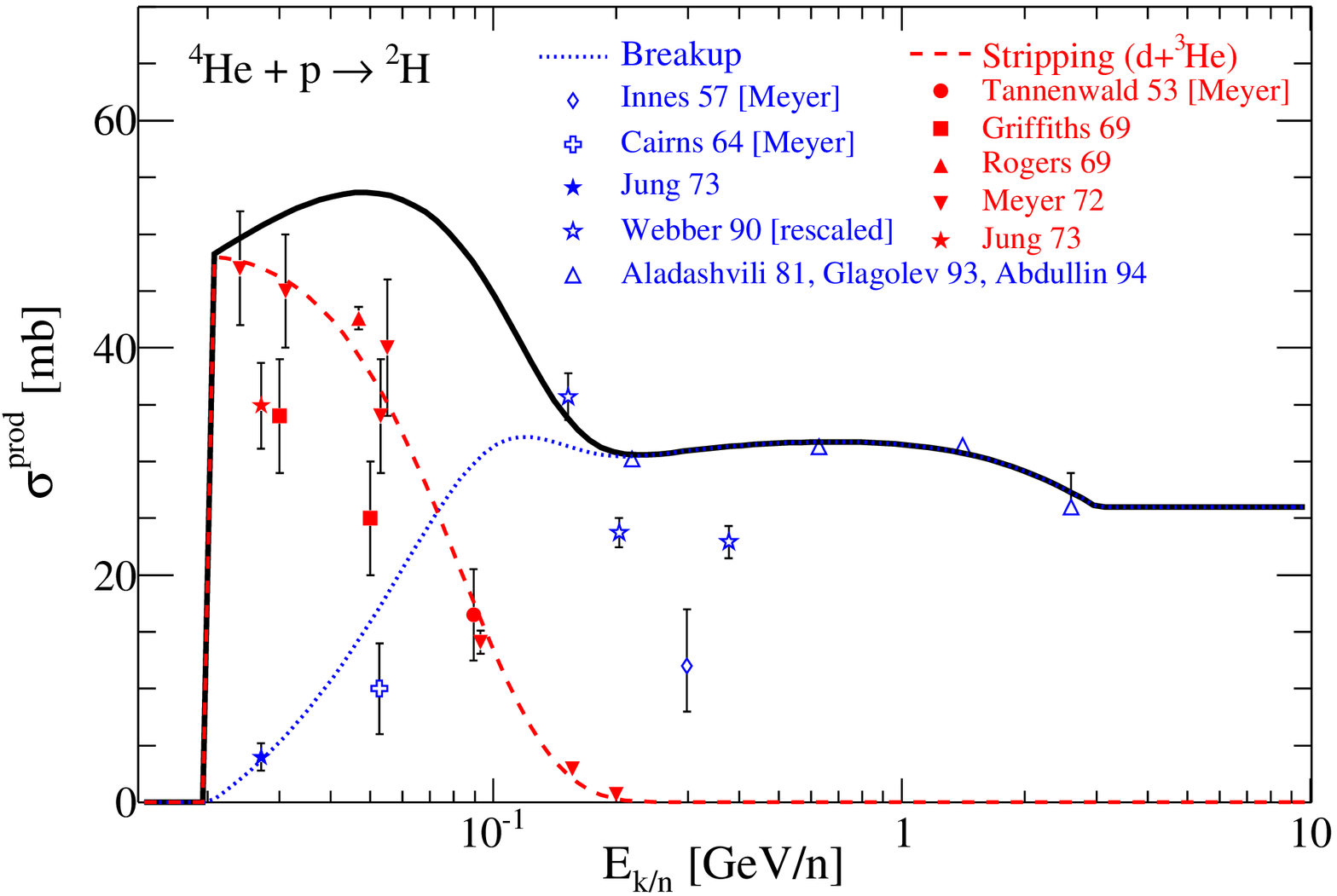}
\includegraphics[width=\columnwidth]{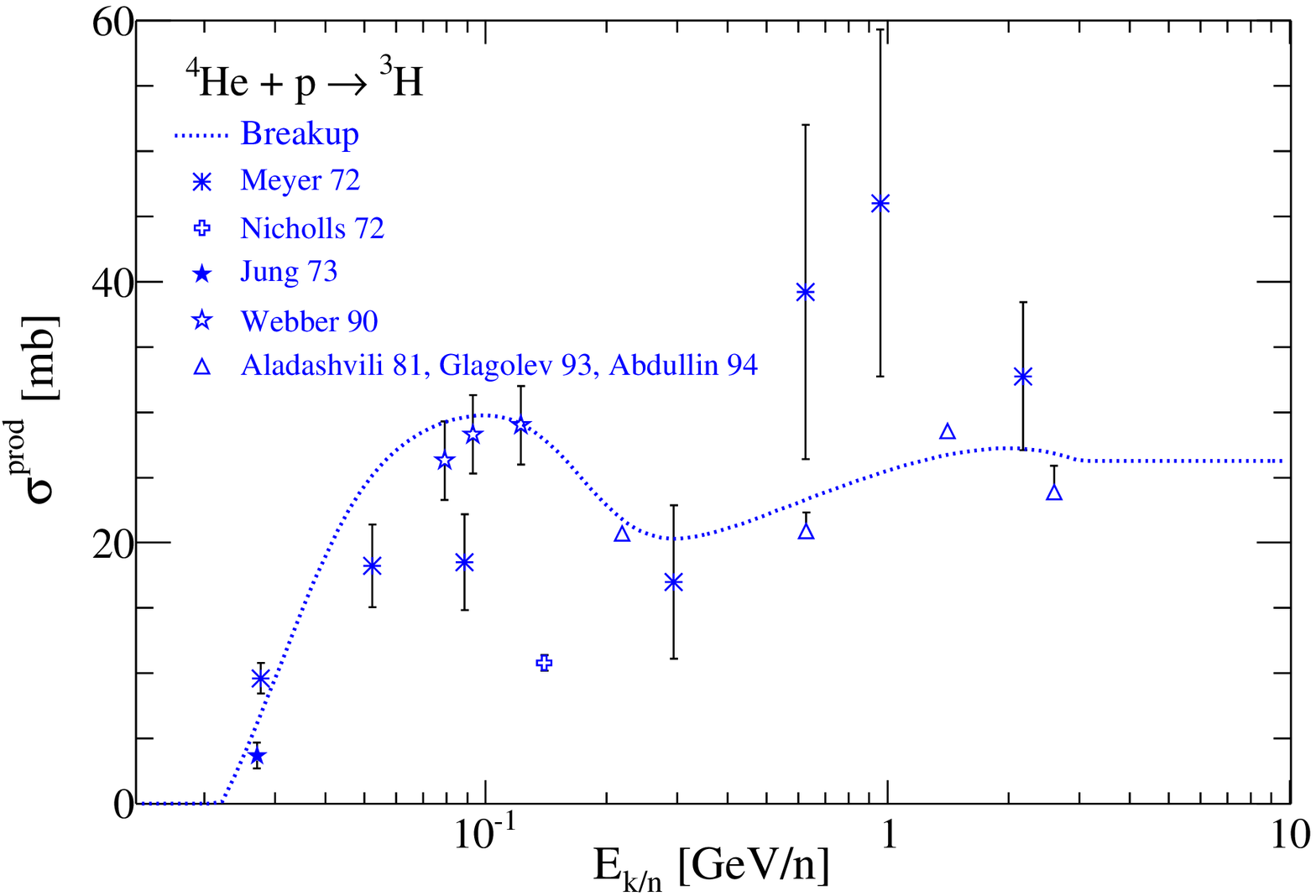}
\caption{Inclusive production cross-sections of \het{} (top), \deut{} (centre) and \trit{} (bottom) in \hef{}+H reaction. The data (see text for details) are from \citet{Tannenwald53,Innes57,Cairns1964,Rogers69,1969ApJ...158..711G,Meyer72,1973PhRvC...7.2209J,Aladashvili81,Webber90,Glagolev:1993cw,1994NuPhA.569..753A}.
}
\label{fig:sigprod_4He_p}
\end{center}
\end{figure}

Recent and illustrative reviews on \hef{}+H reaction and the production of light fragments
is given by \citet{1990NuPhA.516...77B,Cucinotta,Blinov2008}. As said earlier, we are only interested in
the total inclusive production cross-section, not in all the possible numerous final states 
\citep[see, e.g., Table~3 of][]{Blinov2008}. We adapt the parametrisation
of \citet{Cucinotta}, which takes into account separately the
break-up and stripping (for \het{} and \deut{}) cross-sections. The former reaction corresponds
to the case where the helium nucleus breaks up leading to coalescence of free nucleons into a
new nucleus. The latter happens via the pickup reaction where the incident proton tears a
neutron or a proton off the helium nucleus. Both reaction and the total are shown along with
the experimental data in Fig.~\ref{fig:sigprod_4He_p}.

The most accurate set of data (upward blue empty triangles) are from the experiments set up
in ITEP and LHE JINR (\citealt{Aladashvili81,Glagolev:1993cw,1994NuPhA.569..753A}, summarised
in \citealt{Blinov2008}). Their highest energy data point \citep{Glagolev:1993cw} is a
conservative estimate as the more or equal to 6-prong reactions are not detailed (see Table~3
of  \citealt{Blinov2008} and Table~4 of \citealt{Glagolev:1993cw}). To take into account that
possibility, we consider an error of a few mb in the plots of Fig.~\ref{fig:sigprod_4He_p}.
Let us consider in turn each product of interest.

\paragraph{\het{} production}

The stripping cross-section data (d and \het{} in the final state) are well fitted by
Eq.~(130) of \citet{Cucinotta}. However, the \citet{1969ApJ...158..711G} and
\citep{1973PhRvC...7.2209J} are $\sim 30\%$ below the other data. Actually, for the latter
(filled stars) the break-up cross-section is above other data, it may be that the end
products are misreconstructed (in this or the other experiments). Nevertheless, the sum of
the two|which is the one that matters|is consistent in all data. Note that we slightly
modified the break-up cross-section provided by \citet{Cucinotta} to better fit the
high-energy data points. For the latter, all the data are consistent with one another, but
for the high precision ITEP data at 200 MeV/n.

\paragraph{\deut{} production}

The stripping cross-section is as for \het{} (d and \het{} in the final state). The high-energy
break-up cross-section data (LHE JINR and \citealt{Webber90}) are inconsistent. We have decided
to rescale the Webber data, to take into account the fact that in his preliminary account of
the results \citep{Webber90}, the total inelastic cross-section is smaller than that given
in a later and updated study \citep{1997AdSpR..19..755W}. Still, the agreement between the
two sets is not satisfactory. The other high-energy data point is the \citet{Innes57}
experiment, and it suffers large uncertainties and maybe systematics (it is for n $+$ \hef{}
reaction, and the data point is provided by \citet{Meyer72} who relied on several assumptions
to get it). The ITEP/LHE JINR data being the best available, we have replaced the formula for the
\deut{} breakup of \citet{Cucinotta} by a form similar as that given for \het{}, but where we
changed the parameters to fit the high energy points.

\paragraph{\trit{} production}
There is only break-up for the \citet{Cucinotta} \trit{} production. The data are in broad
agreement with one another, but for the \citet{Nicholls1972} point (open plus). Again, we
have adapted the \citet{Cucinotta} parametrisation to better fit the ITEP/LHE JINR data.

\subsubsection{\het{} + p $\rightarrow$ \deut{} (breakup) and p + p 
$\rightarrow$ \deut{} (fusion)}

\begin{figure}[!t]
\begin{center}
\includegraphics[width=\columnwidth]{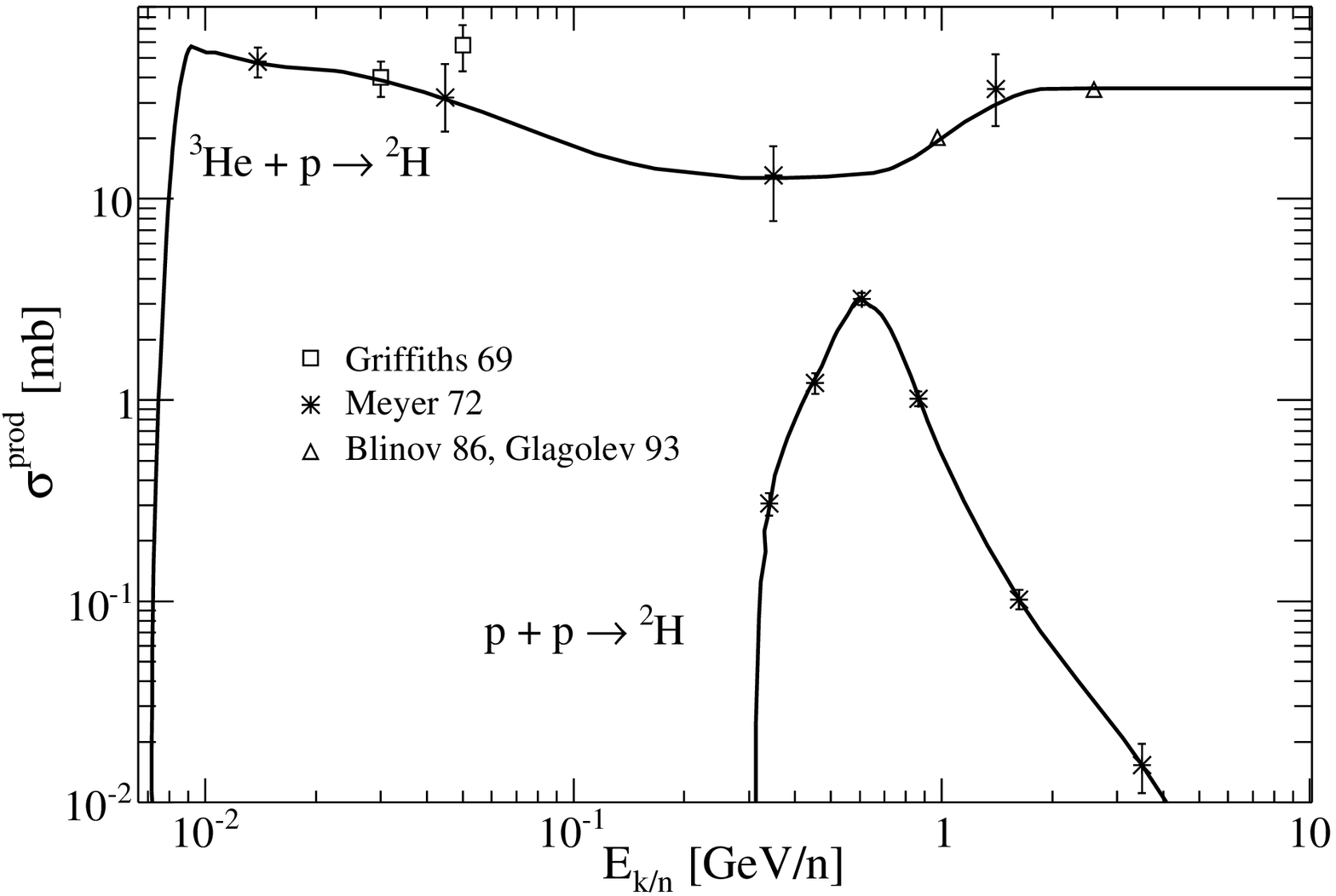}
\caption{\deut{} other production channels from the less abundant \het{} and the peaked fusion pp reaction (the much smaller cross-section is redeemed by a CR flux higher (p instead of \het{}). The data are from \citet{1969ApJ...158..711G,Meyer72,Blinov:1985hj,Glagolev:1993cw}.
}
\label{fig:sigprod2H_other}
\end{center}
\end{figure}

There are two other channels for producing \deut{} from light nucleus reactions,
and they are shown in Fig.~\ref{fig:sigprod2H_other} along with the data. The
first one is from \het{} (break-up and stripping). The CR flux of the latter is
less abundant than the \hef{} flux. With a ratio of $\sim 20\%$ at 1 GeV/n
(decreasing at higher energy) and similar production cross-sections ($\sim
30-40$~mb), this is expected to contribute by the same fraction at GeV/n
energies, and then to become negligible $\gtrsim 10$~GeV/n. The second channel is
the \deut{} coalescence from two protons. The cross-section is non-vanishing only
for a very narrow energy range. Even if the cross-section is 10 times smaller
than for the other channels, the fact that CR protons are $\sim 10$ times more
numerous than \hef{} makes it a significant channel slightly below 1 GeV/n.

The fitting curves are taken from \citet{Meyer72}, but we adapted the fit for
the \het{}+p channel to match the two high-energy ITEP/LHE JINR data points. 

\subsubsection{Proj$_{(A>4)}$ + p $\rightarrow$ \deut{}, \trit{}, and \het{}}
  
For nuclear fragmentation cross-sections of heavier nuclei,
the concepts of `strong' or `weak' factorisation relies on the
fact that at high energy enough, the branching of the various
outgoing particles-production channels becomes
independent of the target. This corresponds to the factorisation
$\sigma^{\rm strong}(P,F,T) = \gamma_{\rm P}^{\rm F} \,\gamma_{\rm T}$ or $\sigma^{\rm weak}(P,F,T) =
\gamma_{\rm P}^{\rm F} \,\gamma_{\rm PT}$ where $\sigma(P,F,T)$ is the
fragmentation cross-section for the projectile P incident upon
the target T producing the fragment F. This is discussed,
e.g., in \citet{1983PhRvC..28.1602O}, where it is concluded that although
strong factorisation is probably violated, weak factorisation
seems exact (see also, e.g., \citealt{1995NIMPB.103..183M}).
The parametrisation proposed below takes advantage of it.

\begin{figure*}[!t]
\begin{center}
\includegraphics[width=\columnwidth]{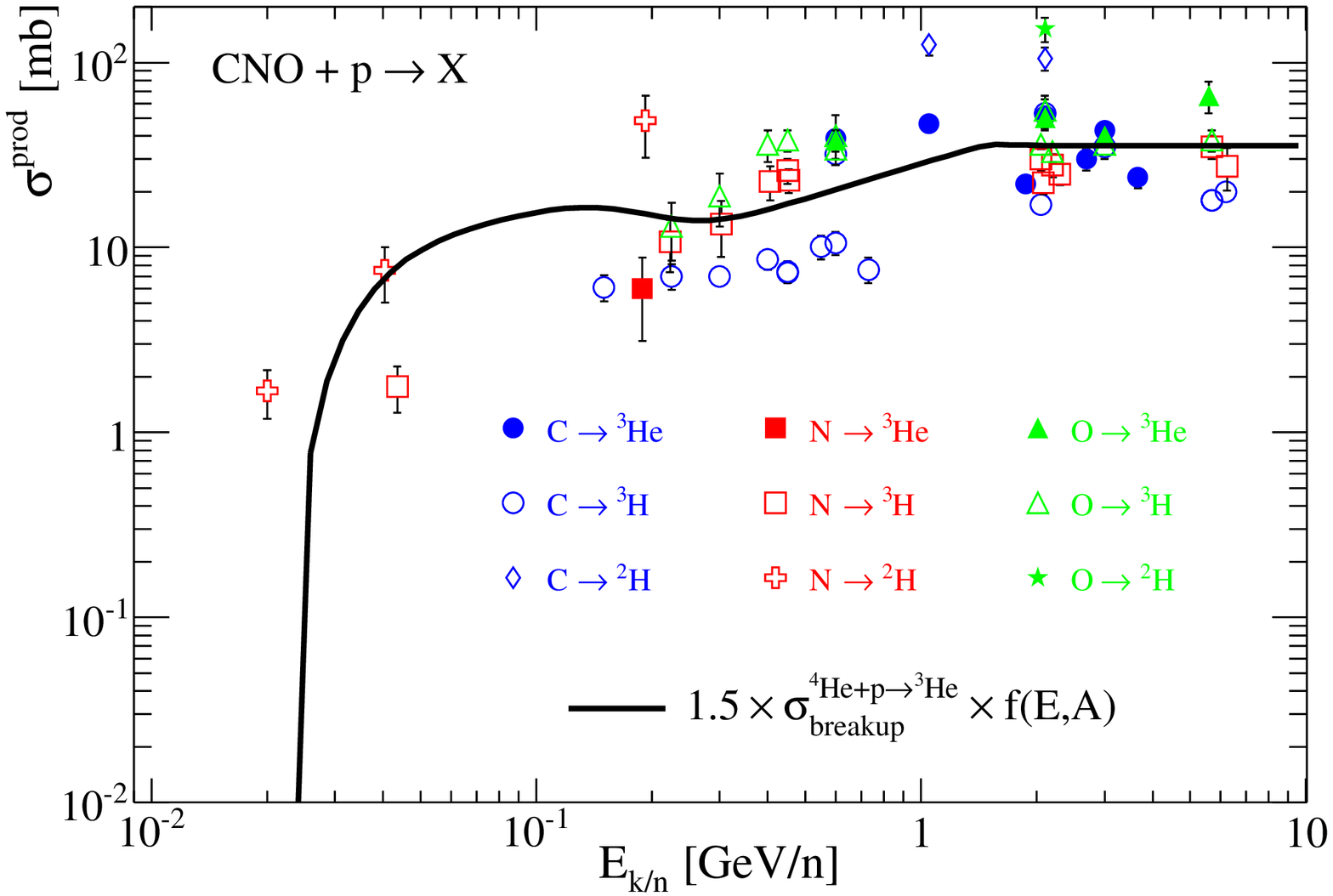}
\includegraphics[width=\columnwidth]{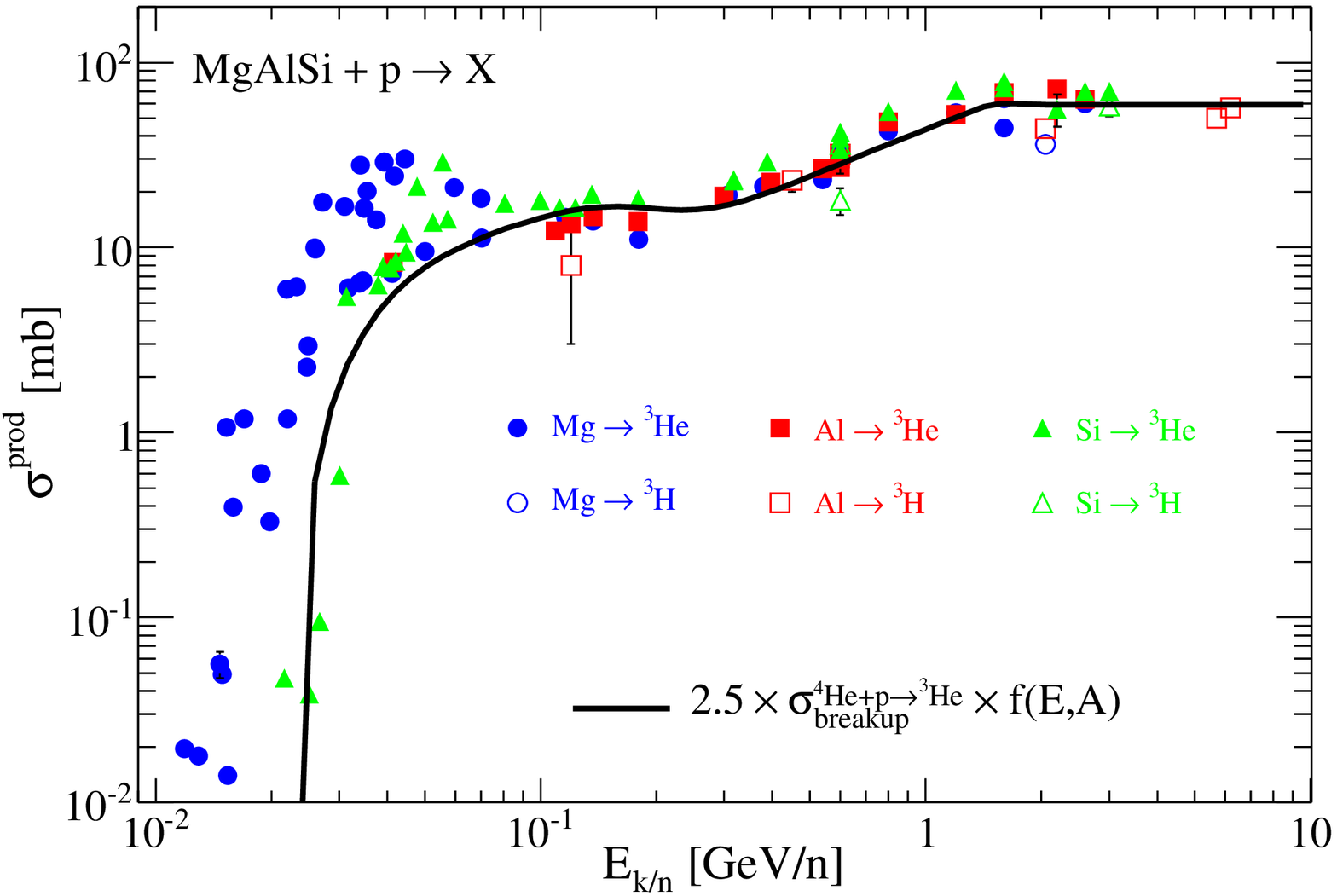}
\includegraphics[width=\columnwidth]{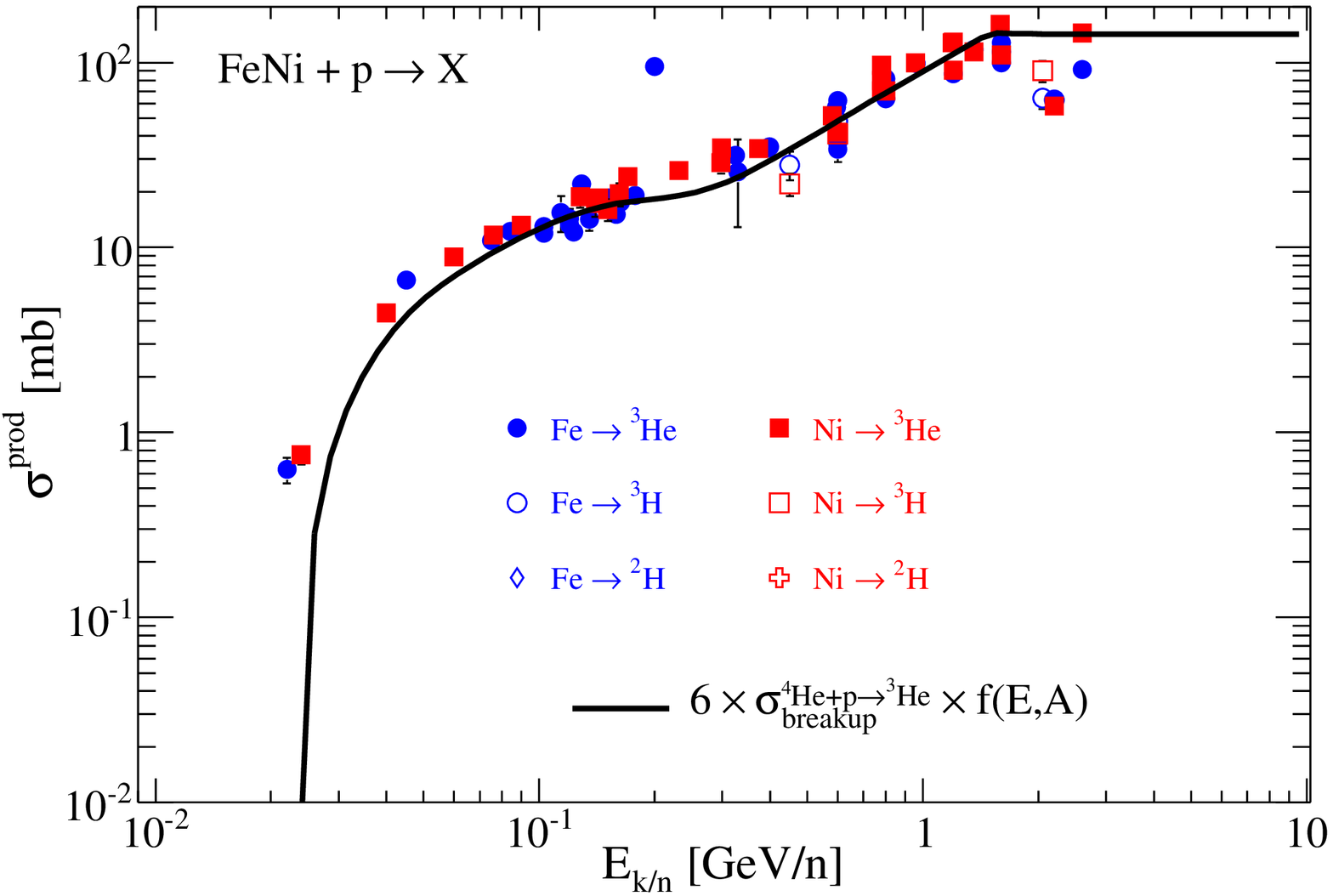}
\includegraphics[width=\columnwidth]{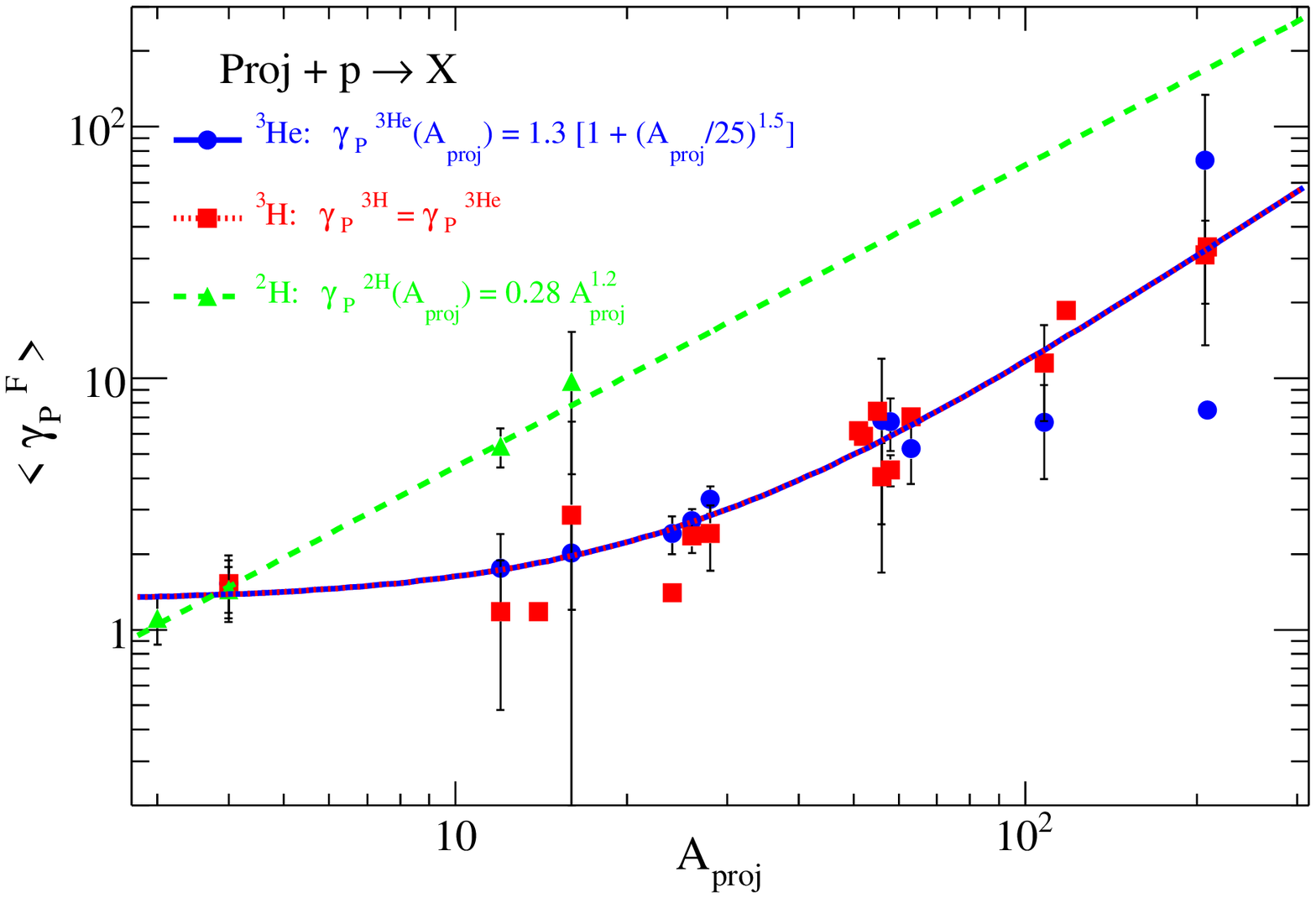}
\caption{Proj$_{(A>4)}$ + p $\rightarrow$ \deut{}, \trit{}, and \het{} cross-sections for Proj=C,N,O (top left),
Proj=Mg,Al,Si (top right), and Proj=Fe,Ni (bottom left). The bottom right correspond to the
$\gamma_P^F$ factor (see Eq.~\ref{eq:Nucp1} and text for explanations). The references for the data
are gathered in Table~\ref{tab:refs_prod}.
}
\label{fig:sigprodNucp}
\end{center}
\end{figure*}

\begin{table*}[!t]
\caption{References for the light nuclei production cross-sections
(references for \hef{} projectiles are given in Fig.~\ref{fig:sigprod_4He_p}).}
\label{tab:refs_prod}
\centering
\begin{tabular}{llccll} \hline\hline
  Proj.      & Frag. & \#data& $E_{k/n}$ & Ref. \\\hline
& \multicolumn{1}{c}{} \vspace{-0.30cm} \\
 N, O, Fe                    &  \trit{}            &    3  &    2.2        & \citet{1955PhRv...97.1303F}   \\
 C, N, O, Mg, Al, Fe, Ni, Ag, Sn, Pb  &  \trit{}            &   26  &  0.45-6.2     & \citet{1959PhRv..114..878C,1956PhRv..101.1557C} \\
    C                        &  \trit{}            &   4   &  0.225-0.73   & \citet{1960PhRv..118.1618H}   \\
    Mg, Al                   &  \het{}           &   2   &      0.54     & \citet{Bieri62}               \\
C,O,Mg,Al,Si,V,Cr,Mn,Fe,Ni,Cu,Ag,Pb,Bi  & \trit{}, \het{}     &  38   &  0.225-5.7    & \citet{Goebel64}   \\
    CNO                      & \deut{}, \trit{}, \het{}& 14  &  0.02-7.5     & \citet{1969ApJ...155..587R}   \\
   C, O, Si                  & \trit{}, \het{}     &  12   &  0.6-3.0      & \citet{1973PhRvC...7.2179K}   \\
    Si, Mg                   & \het{}            &  33   &  0.02-0.06    & \citet{1976JGR....81.5689W}   \\
     Mg                      & \het{}            &   6   &  0.015-0.07   & \citet{Pulfer79}              \\
    C, O                     & \deut{}, \trit{}, \het{} & 8   &   1.05-2.1    & \citet{1983PhRvC..28.1602O}   \\
     Ag                      & \het{}            &   1   &     0.48      & \citet{1984PhRvC..29.1806G}   \\
   Mg, Al, Si                & \het{}            &   3   &     0.6       & \citet{1989NIMPB..42...76M}   \\
   Mg, Al, Si, Fe, Ni        & \het{}            &  21   &   0.8-2.6     & \citet{1995NIMPB.103..183M}   \\
    Mg, Al, Si               & \het{}            &  33   &   0.015-1.6   & \citet{1998NIMPB.145..449L}   \\
      C                      & \het{}            &  3    &  1.87-3.66    & \citet{2000JPhG...26.1171K,2002JPhG...28.1199K}   \\
      Pb                     & \het{}            &  22   &   0.04-2.6    & \citet{2008NIMPB.266.1030L}   \\
    Fe, Ni                   & \het{}            &  53   &   0.022-1.6   & \citet{2008NIMPB.266....2A}   \\
\hline
\end{tabular}
\end{table*}
Several data exist for the production of light isotopes from nuclei $A\ge12$
on H (see Table~\ref{tab:refs_prod}). The most complete sets of data in terms of
energy coverage are for the projectiles C, N, and O $(\langle A\rangle=14)$, the group
Mg, Al, and Si $(\langle A\rangle=26)$, and the group Fe and Ni $(\langle A\rangle=57)$. They are plotted
in Fig.~\ref{fig:sigprodNucp} (top panels and bottom left panel).
The solid lines correspond to an adjustment (by eye), rescaled from
the $\sigma^{\hef{}{\rm p}\rightarrow \het{}}_{\rm breakup}$ cross-section
(because heavy projectile do not give $A=3$ fragments in the stripping process).
As $\sigma^{\rm prod}_{\het{}} \approx \sigma^{\rm prod}_{\trit{}}$, no distinction is made
for the fit (\deut{} data are scarce and do not influence the conclusions
drawn from these three groups of nuclei). The following parametrisation
 \begin{equation}
    \sigma^{\rm Pp \rightarrow \rm F} (E_{k/n}, A_{\rm P}) = \gamma_{\rm P}^{\rm F} \cdot f(E_{k/n}, A_{\rm P}) \cdot 
      \sigma^{\hef{}{\rm p}\rightarrow \het{}}_{\rm breakup}(E_{k/n})\;,
  \label{eq:Nucp1}
 \end{equation}
with 
\begin{equation}
  f(E_{k/n}, A_{\rm P}) \!=\! 
    \begin{cases}
      \left(\frac{E_{k/n}}{1.5~{\rm GeV/n}}\right)^{0.8 \cdot \sqrt{\frac{A_{\rm P}}{26}}} \!\!\!& \text{if $E_{k/n}\!<\!1.5$ GeV/n,}\\
       ~1 & \text{otherwise;}
\end{cases}
  \label{eq:Nucp2}
\end{equation}
proves to fit well the three groups of data for energies greater than a few tens of MeV/n.
Thanks to the $f(E_{k/n}, A_{\rm proj})$ factor, there is no further energy dependence in
the $\gamma_{\rm P}^{\rm F}$ factor, so that the latter can be determined from the data points at any energy.
The bottom right panel of Fig.~\ref{fig:sigprodNucp} shows the measured mean value and
dispersion\footnote{In practice, for a given $A$,
the mean and dispersion are calculated from all the existing point above a projectile energy
150 MeV/n.} as a function of $A$, from which we obtain:
 \begin{eqnarray}
    \gamma_{\rm P}^{\het{}} &=& \gamma_{\rm P}^{\trit{}} = 1.3\, \left[ 1 + \left( \frac{A_{\rm P}}{25}\right)^{1.5}\right], \nonumber\\
    \gamma_{\rm P}^{\deut{}} &=& 0.28\, A_{\rm P}^{1.2}.
  \label{eq:Nucp3}
 \end{eqnarray}
The set of formulae (\ref{eq:Nucp1}), (\ref{eq:Nucp2}), and (\ref{eq:Nucp3})
completely define the Proj+p production cross-sections for the light fragments.

\subsection{Proj$_{(A\ge4)}$ + \hef{} $\rightarrow$ \deut{}, \trit{}, and \het{}}
Data for T + \hef{} where the target T is heavier than p
are scarce. In a compilation of \citet{1995ICRC....2..622D},
the authors find that the \het{} production scales as $A_{\rm T}^{0.31}$
(based on 4 data point with $A_{\rm T}\ge 7$). This is the scaling we employ for the \trit{} and \deut{} production as well.

\bibliographystyle{aa}
\bibliography{quartet}
\end{document}